\documentclass[
reprint,
amsmath,
superscriptaddress,
noeprint,
prl,
]{revtex4-1}
\usepackage{siunitx}
\usepackage{graphicx}
\usepackage{dcolumn}
\usepackage{braket}
\usepackage{xfrac}
\usepackage{hyperref}
\usepackage[usenames,dvipsnames]{color}
\usepackage{tikz-cd}
\usepackage{float}
\usepackage[nameinlink,capitalise]{cleveref}
\usepackage{mathtools}
\usepackage{amsmath}
\usepackage{bm}
\DeclareSIUnit\gauss{G}
\DeclareSIUnit{\au}{{a.u.}}
\hypersetup{
colorlinks=true,
linkcolor=blue,
citecolor=blue,
filecolor=green,
urlcolor=blue,
}

\begin{document}
\title{Stereodynamic control of overlapping resonances in cold molecular collisions}

\author{Masato Morita}
\affiliation{Department of Chemistry and Biochemistry, University of Nevada, Las Vegas, Nevada 89154, USA}
\author{Qian Yao}
\affiliation{Department of Chemistry and Chemical Biology, University of New Mexico, Albuquerque, New Mexico 87131, USA}
\author{Changjian Xie}
\affiliation{Department of Chemistry and Chemical Biology, University of New Mexico, Albuquerque, New Mexico 87131, USA}
\affiliation{Institute of Modern Physics, Shaanxi Key Laboratory for Theoretical Physics Frontiers, Northwest University, Xian, Shaanxi 710127, China}
\author{Hua Guo}
\affiliation{Department of Chemistry and Chemical Biology, University of New Mexico, Albuquerque, New Mexico 87131, USA}
\author{Naduvalath Balakrishnan}
\affiliation{Department of Chemistry and Biochemistry, University of Nevada, Las Vegas, Nevada 89154, USA}

\date{\today}
\begin{abstract}
Stereodynamic control of resonant molecular collisions has emerged as a new frontier in cold molecule research. Recent experimental studies have focused on weakly interacting molecular systems such as HD collisions with H$_2$, D$_2$ and He. We report here the possibility of such control in strongly interacting systems taking rotational relaxation in cold collisions of HCl and H$_2$. Using explicit quantum scattering calculations in full six dimensions it is shown that robust control of the collision dynamics is possible even when multiple (overlapping) shape-resonances coexist in a narrow energy range, indicating that cold stereochemistry offers  great promise for many  molecules beyond simple systems. We demonstrate a striking case where two prominent peaks in overlapping resonances are switched-off simultaneously by suitable alignment of the HCl molecule. 
\end{abstract}

\maketitle


Recent progress  in cooling and trapping molecules below 1 Kelvin has led to new insights into chemical dynamics, quantum information processing, quantum simulation, and the test of fundamental physics beyond the standard model \cite{Rev_Julienne06,Rev_Krems08,Rev_Carr08,Rev_Costes_2014,Rev_Ye2014,Rev_Bala16,Rev_Bohn17,Rev_Safronova18,Merge,Superconduct}. In particular, the ability to prepare molecules in selected motional, internal and orientational quantum states has allowed interrogation of molecular events with unprecedented precision. The suppression of higher-order angular momentum partial waves in cold collisions provides an ideal platform to explore state-to-state chemistry with partial wave resolution. 

A series of papers by Perreault {\it et al}. \cite{2017_Science_Perreault,2018_NatChem_Perreault,SARP_HD-He} laid groundwork for cold stereodynamics \cite{Zare_Rev,Orr-Ewing,Aldegunde,Kopin} by combining 
cold molecular beam technology and coherent optical control of molecular states. 
They demonstrated stereodynamic control of rotational quenching of vibrationally excited HD ($v=1,j=2$) in cold collisions with  H$_2$, D$_2$ and He  \cite{2017_Science_Perreault,2018_NatChem_Perreault,SARP_HD-He} in a single molecular beam.  The initial molecular state of HD, including the bond axis alignment, was prepared by Stark-induced adiabatic Raman passage (SARP) \cite{SARP_1,SARP_2}. 
The quenching rate and angular distribution were found to vary significantly depending on whether the HD bond axis is preferentially aligned horizontal (H-SARP) or vertical (V-SARP) to the initial relative velocity vector between the collision partners, providing conclusive evidence of strong stereodynamic effect in HD+H$_2$/D$_2$ collisions near 1 Kelvin.

While the collision energies involved in the experiment are narrow ($<5$ K) and only a limited number of partial waves ($\sim l< 5$) is involved, the experiments do not provide energy or partial wave resolution directly. In a subsequent theoretical study Croft {\it et al}. \cite{2018_PRL_Croft} attributed key feature of the angular distribution of the rotationally de-excited HD to an incoming {\it d\,}-wave ($l=2$) shape resonance centered at about 1 K. 
The role of isolated (shape) resonances in controlling quenching rate and overall angular distribution of the products was further explored in  theoretical  studies of  HD+H$_2$ \cite{2019_JCP_Croft,2019_PRL_Jambrina} and in recent experiments on HD+He \cite{SARP_HD-He}.

As a straightforward extension of H-SARP and V-SARP, it is possible to consider a continuous control of the preferential alignment of the HD bond axis to the initial relative velocity vector  by varying the polarization direction of the SARP laser  with respect to molecular beam axis.
This allows explicit control of the scattering resonance as demonstrated by theoretical studies of Croft and Balakrishnan \cite{2019_JCP_Croft} and Jambrina {\it et al} \cite{2019_PRL_Jambrina} for HD+H$_2$. 
We note that, experimental SARP technology has the capability to achieve such flexible control of molecular states \cite{SARP_HD-He,SARP_2} and it is applicable to a broad class of molecules including those with no permanent electric dipole moment \cite{SARP_H2,SARP_D2}. 

These advances now allow controlled studies of molecular collisions and chemical reactions using fully tailored molecular initial rovibrational states ($v, j, m$) and their coherent superpositions, where $m$ is the projection of $j$ onto a quantization axis. 
Such investigations of systems beyond HD+H$_2$, HD+D$_2$ and HD+He are crucial to unveil whether the stereodynamic control is possible for broader classes of molecules with stronger intermolecular interactions. In particular, stereodynamic effect in the presence of multiple (overlapping) resonances in cold collisions is an open question that remains to be addressed. 

In this Letter, we examine this question by taking  rotational quenching of HCl by collisions with H$_2$ in the sub-Kelvin regime based on full-dimensional quantum scattering calculations on a recently reported {\it ab initio} potential energy surface (PES) \cite{H2HCl_PES}. This system is of current  experimental interest in Zare's group (private  communication).
In comparison with HD, the HCl molecule is strongly dipolar (dipole moment $\sim$ 1.08 D) 
and is amenable to high optical controllability of its internal states. 
The HCl+H$_2$ interaction potential is deeper and more anisotropic  \cite{H2HCl_PES} leading to strong coupling between the different rovibrational channels at short range. 
The small rotational constant  of HCl (a factor of 5 smaller than HD) also contributes to the increased density of states and more complex collision dynamics for the HCl+H$_2$ system.


Details of the {\it ab initio} PES for HCl+H$_2$ and its analytical representation are reported previously \cite{H2HCl_PES}. 
Briefly, the PES consists of a short-range part fit to high-level ab initio data and a long-range part giving an accurate description of the electrostatic and dispersion interactions. The depth of the interaction potential well is $\sim$ 216 cm$^{-1}$, significantly deeper than that between two H$_2$ molecules ($\sim$ 35 cm$^{-1}$).
The time-independent Schr\"{o}dinger equation for the molecular scattering is numerically solved using the close-coupling method as implemented in the TwoBC code \cite{TwoBC,CC_3}. More details of the computational approach can be found in previous publications  \cite{2018_PRL_Croft,2019_JCP_Croft,H2HCl_PES} and we limit our discussions to key notations and essential formulas. 

The state-to-state differential cross section (DCS) for isotropic collisions (no alignment) in the helicity representation is given by \cite{DCS}
\begin{equation}\label{eq:DCS_k}
\small
\begin{split}
     \frac{ d\sigma_{{\alpha} \to \alpha'} } {d\Omega} 
    = \frac{1}{(2j_1+1)(2j_2+1)} 
     \sum^{}_{k_1,k_2,k'_1,k'_2}
  |q_{\alpha,k_1,k_2 \to 
      \alpha', k'_1,k'_2}|^2,
\end{split}
\normalsize
\end{equation}
where $\alpha\equiv v_1j_1v_2j_2$ and $\alpha'\equiv v'_1j'_1v'_2j'_2$ refer to the initial and final combined molecular rovibrational states, respectively, $d\Omega$ is the infinitesimal solid angle,
the subscript 1 refers to HCl and 2 to H$_\text{2}$. The quantum number $k$/$k'$ is the projection of the initial/final molecular rotation angular momentum $j/j'$ onto the BF z-axis. 
The $\theta$ dependence in the scattering amplitude is given as \cite{DCS,2018_PRL_Croft,2019_JCP_Croft}
\begin{equation}\label{eq:q_k}
\small
\begin{split}
& q_{\alpha,k_1,k_2  \to  \alpha', k'_1,k'_2} \\
&= \frac{1}{2 k_{\alpha}} 
      \sum^{}_{J} (2J+1)  
 \sum^{}_{j_{12}, j'_{12}, l, l'} i^{l-l'+1}  T^{J}_{\alpha j_{12} l,\alpha' j'_{12} l'}(E) d^{J}_{k_{12},k'_{12}}(\theta) \\
&\ \ \ \  \times <j'_{12} k'_{12} J -k'_{12}|l' 0> <j_{12} k_{12} J -k_{12}|l 0> \\
&\ \ \ \  \times <j'_{1} k'_{1}   j'_{2} k'_{2}|j'_{12} k'_{12}> <j_{1} k_{1} j_{2} k_{2}|j_{12} k_{12}>,
\end{split}
\normalsize
\end{equation}
where $k_{\alpha}$ is the wave number for the incident channel, $J$ is the quantum number of the total angular momentum of the collision complex defined as $\bm{J}={\bm j}_{12}+{\bm l}$ with the combined rotational angular momentum ${\bm j}_{12}$ of the two molecules given by ${\bm j}_{12}={\bm j}_1+{\bm j}_2$ and the relative orbital angular momentum ${\bm l}$ between the collision partners. The $T$-matrix is given in terms of the $S$-matrix as $T=1-S$, $E$ is the total energy, and $d^{J}_{k_{12},k'_{12}}(\theta)$ is the Wigner's reduced rotation matrix, and the braket $< \ | >$ denotes a Clebsch-Gordan coefficient. We note that $k_{12}$ and $k'_{12}$ in \cref{eq:q_k} mean $k_1+k_2$ and $k'_1+k'_2$, respectively.

By taking the integral of the DCS in \cref{eq:DCS_k} over $\theta$ and $\phi$ one obtains the state-to-state integral cross section (ICS) as \cite{DCS,CC_3,H2HCl_PES}
%
\begin{equation}\label{eq:xs}
\small
\begin{split}
    & \sigma_{{\alpha} \to \alpha'} (E) 
    =\frac{\pi}{(2j_1+1)(2j_2+1)k_{\alpha}^2} \\
    & \times \sum^{}_{j_{12}j'_{12},ll'J}(2J+1) 
  |\delta_{\alpha j_{12} l,\alpha' j'_{12} l'}-S^{J}_{\alpha j_{12} l,\alpha' j'_{12} l'}(E)|^2.
\end{split}
\normalsize
\end{equation}

\cref{eq:DCS_k,eq:xs} include averaging over all possible initial orientations of the molecular rotational states specified by the projections $k_1$ and $k_2$.  
Our primary goal is to explore the effect of aligning the HCl bond axis relative to the initial relative velocity vector for the collision (BF z-axis),
as explored in  previous studies of HD \cite{2017_Science_Perreault,2018_NatChem_Perreault,2019_JCP_Croft,2019_PRL_Jambrina}. 
In this scheme the initial rotational state of HCl is prepared as  $|j_1,\tilde{m}_1=0>$ 
by the laboratory fixed SARP laser where $\tilde{m}_1$ 
is the projection of $j_1$ onto the linear polarization direction of the SARP laser.
The prepared state $|j_1,\tilde{m}_1=0>$ 
can be expressed as a coherent superposition of a set of states $|j_1, k_1>$  specified by the projection $j_1$ onto the BF z-axis as
as \cite{2019_JCP_Croft} 
\begin{equation}\label{eq:align}
\small
\begin{split}
|j_1,\tilde{m}_1=0>=\sum^{j_1}_{k_1=-j_1} d^{j_1}_{\tilde{m}_1=0,k_1}(\beta)\ |j_1,k_1>,
\end{split}
\normalsize
\end{equation}
where $\beta$ is the angle between the aligned HCl bond axis and the initial relative velocity vector for the collision. 

For rotational quenching of aligned HCl by cold collision with {\it para\,}-H$_2$ ($j_2=0$, $k_2=0$), the state-to-state DCS with respect to the scattering angle $\theta$ is given by \cite{2019_JCP_Croft}
\begin{equation}\label{eq:SARP_DCS}
\small
\begin{split}
\left(\frac{d\sigma}{d\theta}\right)_\beta=
2\pi \text{sin}\theta \sum^{j_1}_{k_1=-j_1} \sum^{j'_1}_{k'_1=-j'_1} |d^{j_1}_{0,k_1}(\beta)|^2 |q_{\alpha,k_1,0 \to \alpha',k'_1,0}|^2
\end{split}
\normalsize
\end{equation}
where it is assumed that, like in previous experiments \cite{2017_Science_Perreault,2018_NatChem_Perreault,SARP_HD-He}, the rotationally de-excited HCl molecules in a specific rotational state ($j'_1$) are detected with no resolution of their $k'_1$ or
azimuthal angle $\phi$.
The overall effect of the alignment of the HCl molecules is captured in the weight factor $|d^{j_1}_{0,k_1}(\beta)|^2$ for each $|q|^2$ term associated with an initial $k_1$. The behavior of $|d^{j_1}_{0,k_1}(\beta)|^2$ as a function of $\beta$ for $j_1=2$ is displayed in Ref. \cite{2019_JCP_Croft} (see also Fig.~S1 in SM  \cite{SM}).



\begin{figure}[t!]
\begin{center}
\includegraphics[width=\linewidth]{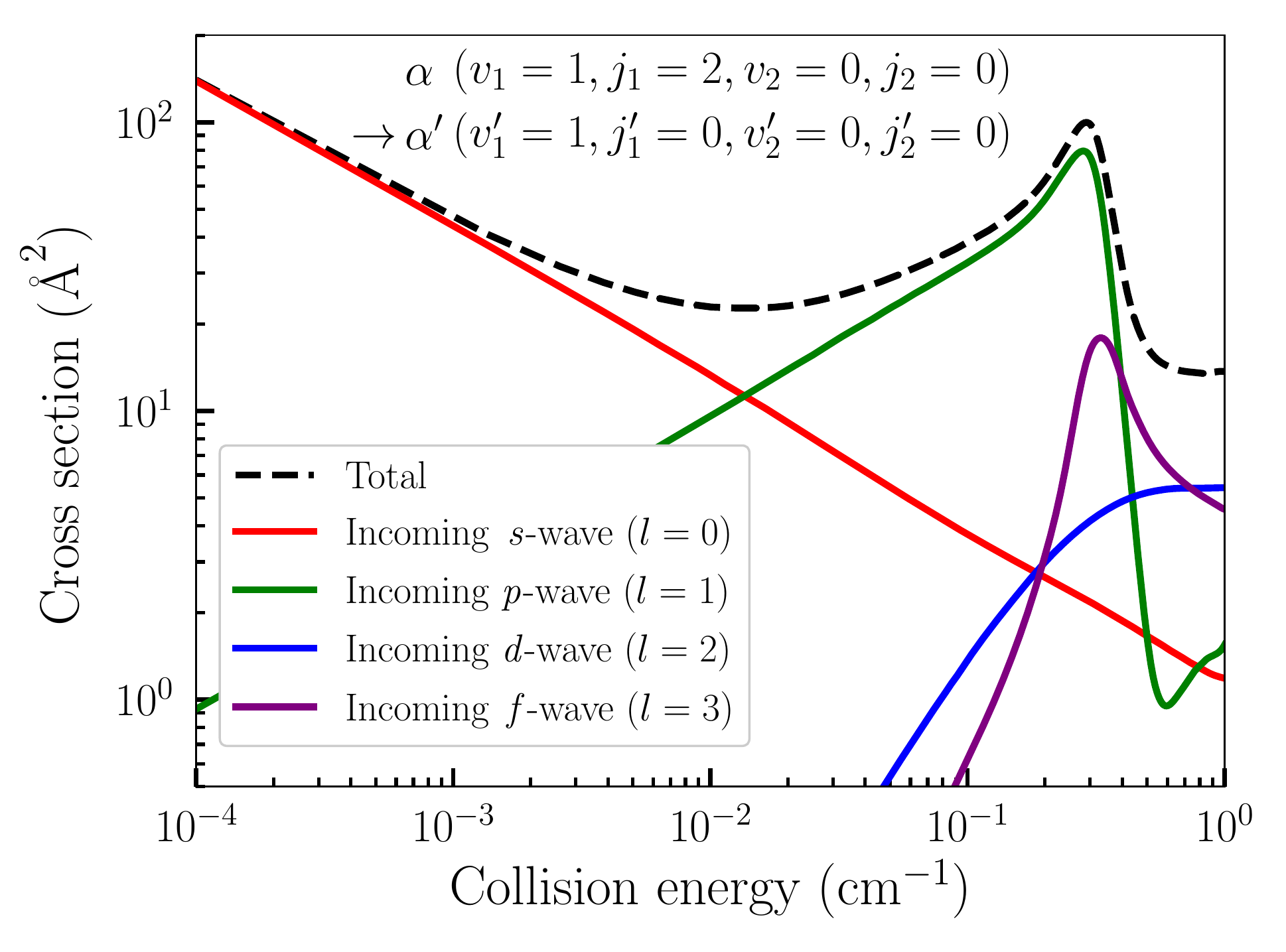}
\end{center}
\caption{Integral cross section for the rotational quenching of HCl ($v_1=1$,\ $j_1=2$ $\rightarrow$ $v'_1=1$,\ $j'_1=0$) due to collisions with {\it para\,}-$\text{H}_2$ ($v_2=0$,\ $j_2=0$) for the isotropic case (no alignment).
The dashed black curve shows the total cross section and the solid curves display  the partial wave contributions.
}
\label{fig_I12F10_ICS}
\end{figure}

\begin{figure}[th!]
\begin{center}
\includegraphics[width=\linewidth]{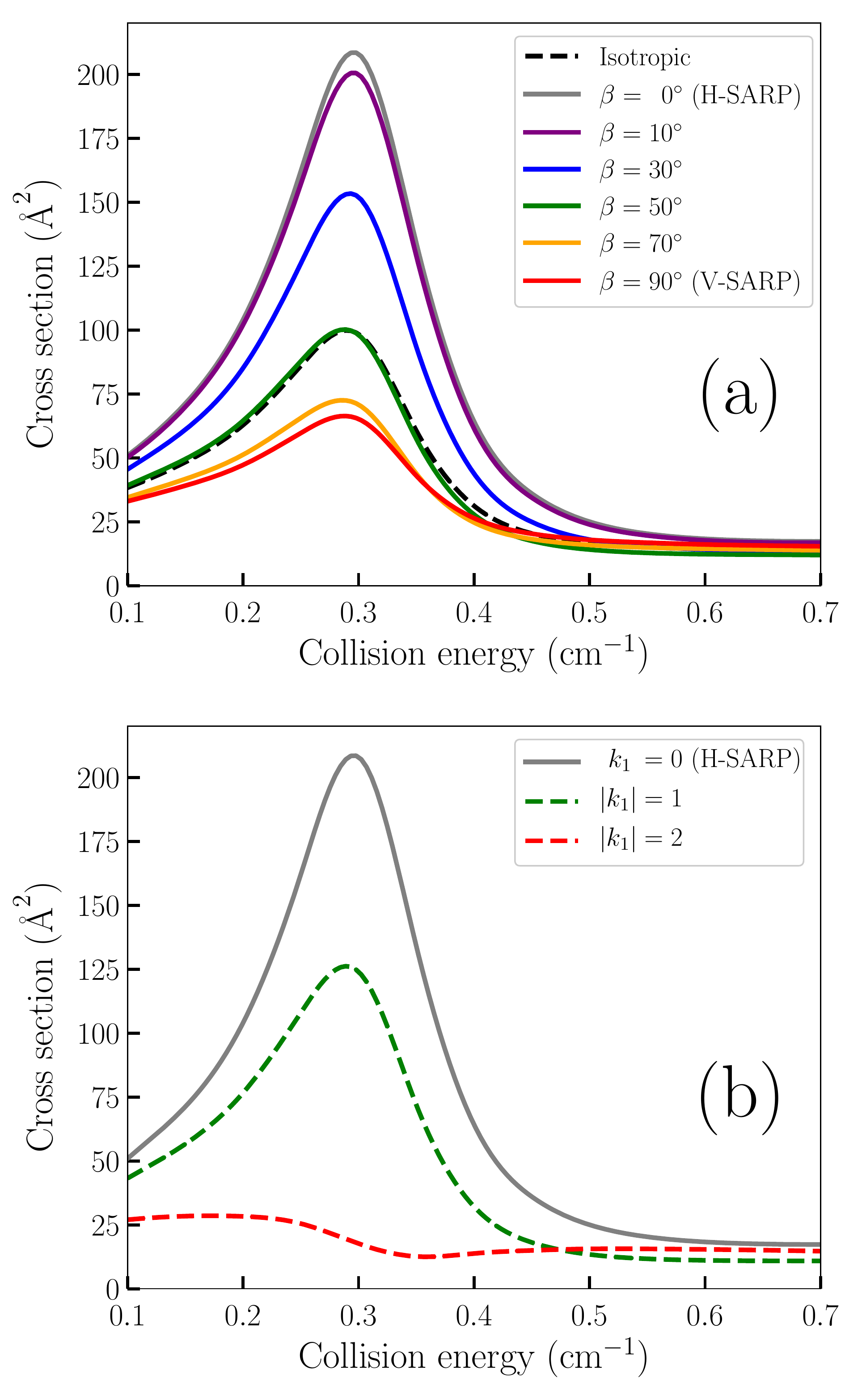}
\end{center}
\caption{
Integral cross section for $\Delta j=-2$ transition in HCl: (a) for different alignments of HCl bond axis against the initial relative velocity vector for the collision.
(b) for different  orientations specified by the projection $k_1$ of the initial rotational state of HCl ($j_1=2$). The result for $k_1=0$ is identical to that of  $\beta=0 ^{\circ}$ in (a). 
}
\label{fig_I12F10_steric}
\end{figure}

First we discuss rotational quenching of HCl ($v_1=1, j_1=2) \to (v'_1=1, j'_1=0$) by collisions with {\it para\,}-H$_2$ ($j_2=0$, $k_2=0$), namely $\Delta j_1=-2$. 
The ICS for isotropic collisions is shown by the dashed curve in \cref{fig_I12F10_ICS}. Contributions from different incoming partial waves $l=0$ to $l=3$ are also shown. It is seen that a pronounced peak at about $E_C=0.3$ cm$^{-1}$ has the main contribution (77 $\%$) from an incoming {\it p\,}-wave ($l=1$)  and a smaller 
contribution from an incoming {\it f\,}-wave ($l=3$). 
The quasi-bound states due to closed channels are inaccessible at these collision energies (see Fig.~S2 in SM \cite{SM}), thus the observed resonance is a {\it p\,}-wave shape resonance.
We note that a {\it p\,}-wave shape resonance was not observed for HD+{\it para\,}-H$_2$ collisions \cite{2018_PRL_Croft,2019_PRL_Jambrina,2019_JCP_Croft}, indicating that cold HCl+H$_2$ collisions exhibit  markedly different features compared to  HD+H$_2$. Indeed, the adiabatic potential energy curves for HCl+H$_2$ are much more complicated \cite{SM} than those of HD+H$_2$ \cite{2018_PRL_Croft}.

We now discuss how to control this resonance by the preparation of initial alignment of the HCl bond axis. 
The ICS in the resonance region with and without initial alignment of HCl is shown in \cref{fig_I12F10_steric} (a) using linear axis scale. The dashed curve (isotropic) is identical to the dashed curve (Total) in \cref{fig_I12F10_ICS}. Other solid curves correspond to different alignment angles of the HCl bond axis with respect to the initial relative velocity vector for the collision, obtained by integrating the corresponding DCS over $\theta$ from 0 to $\pi$ in \cref{eq:SARP_DCS}. The cross section varies almost monotonically by a factor of 3 as $\beta$ increases from 0$^{\circ}$ to 90$^{\circ}$, indicating the feasibility of robust control of quenching rate by changing the angle of alignment.

As shown in \cref{eq:SARP_DCS}, the DCS and the resultant ICS are given as the sum of terms specified by the different projection component $k_1$ of $j_1$. Therefore, the observed trend of the ICS in terms of $\beta$ can be effectively analyzed by examining the ICS with initially oriented HCl specified by a projection $k_1$ (\cref{fig_I12F10_steric} (b)). 
Note that the result for $k_1=0$ is the same as $\beta=0$ in \cref{fig_I12F10_steric} (a).
The cross sections for other values of $k_1$ are relatively suppressed in the resonance region. In particular, a shallow dip is observed for $|k_1|=2$.

The above findings indicate that the sensitive dependence of the dynamics in the resonance region to the initial orientation (specified by $k_1$) of HCl can cause dramatic change in collision outcome. 
Compared to $|k_1|=2$, the ICS for $k_1=0$ is about 12-fold larger at the peak of the resonance ($E_C=0.3$ cm$^{-1}$). Since the projections of the final rotational states are $k'_1=0$ and $k'_2=k_2=0$ due to $j'_1=0$ and $j'_2=j_2=0$, collisions with $k_1=0$ conserve helicity as $k_1+k_2=k'_1+k'_2=0$ and the resonance enhancement can be attributed to a helicity conserving transition. It is worth pointing out that results in \cref{fig_I12F10_steric} (a) do not reflect the full potential of stereodynamic control. If it is possible to specifically prepare an initial HCl rotational state with $k_1=2$ or $k_1=-2$, or their linear combination, the resonance can be completely switched off. Recent advances in  SARP techniques have made progress in this direction \cite{SARP_2, SARP_HD-He}. Indeed, preparation of a phase-locked superposition of $k_1=\pm 1$ states, $\sqrt{1/2}(|j_1=2,k_1=1\ >+\ |j_1=2,k_1=-1>)$, termed X-SARP, was demonstrated for HD \cite{SARP_HD-He}, expanding the range of control achievable using SARP techniques. The range of control expands into the non-resonant region for the DCS as shown in the Supplemental Material \cite{SM}.


\begin{figure}[t!]
\begin{center}
\includegraphics[width=\linewidth]{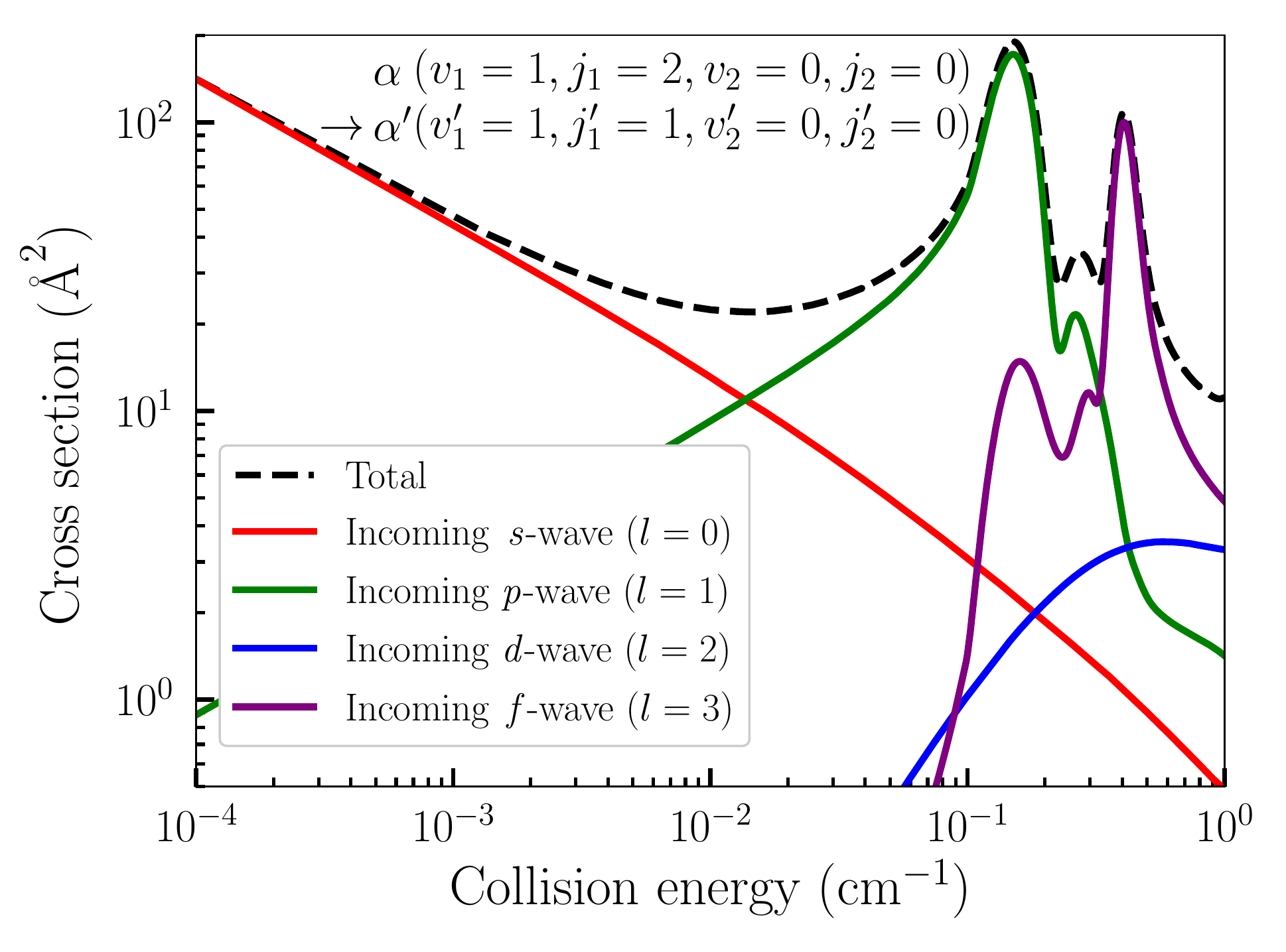}
\end{center}
\caption{
The same as in \Cref{fig_I12F10_ICS} but for $\Delta j=-1$ transition. }
\label{fig_I12F11_ICS}
\end{figure}

Next we examine $\Delta j_1=-1$ transition in HCl, i.e., $(v_1=1$,\ $j_1=2$) $\to$ ($v'_1=1$,\ $j'_1=1$). 
\Cref{fig_I12F11_ICS} presents the total and partial wave resolved ICSs for this process for the isotropic case. 
The crucial difference here in comparison with \cref{fig_I12F10_ICS} is the enhanced incoming {\it f\,}-wave contribution at around 0.4 cm$^{-1}$ in addition to a strong {\it p\,}-wave resonance at about 0.15 cm$^{-1}$ 
leading to  
two prominent peaks. While the two resonances appear as distinct peaks in close proximity we  note that the  {\it p\,}-wave peak has a small contribution from the {\it f\,}-wave resonance illustrating the complex nature of the dynamics and partial overlap of these resonances.
This brings up an important question that has not been addressed  before: can stereodynamic control be effective for such overlapping resonances associated with multiple partial waves? In previous studies of HD+H$_2$, HD+D$_2$ and HD+He, the stereodynamic effect was attributed to the characteristics of a specific  isolated shape resonance. However, such simplicity is not necessarily expected for vast majority of systems due to the higher density of states caused by the stronger interaction and/or heavier mass.



\begin{figure}[t!]
\begin{center}
\includegraphics[width=\linewidth]{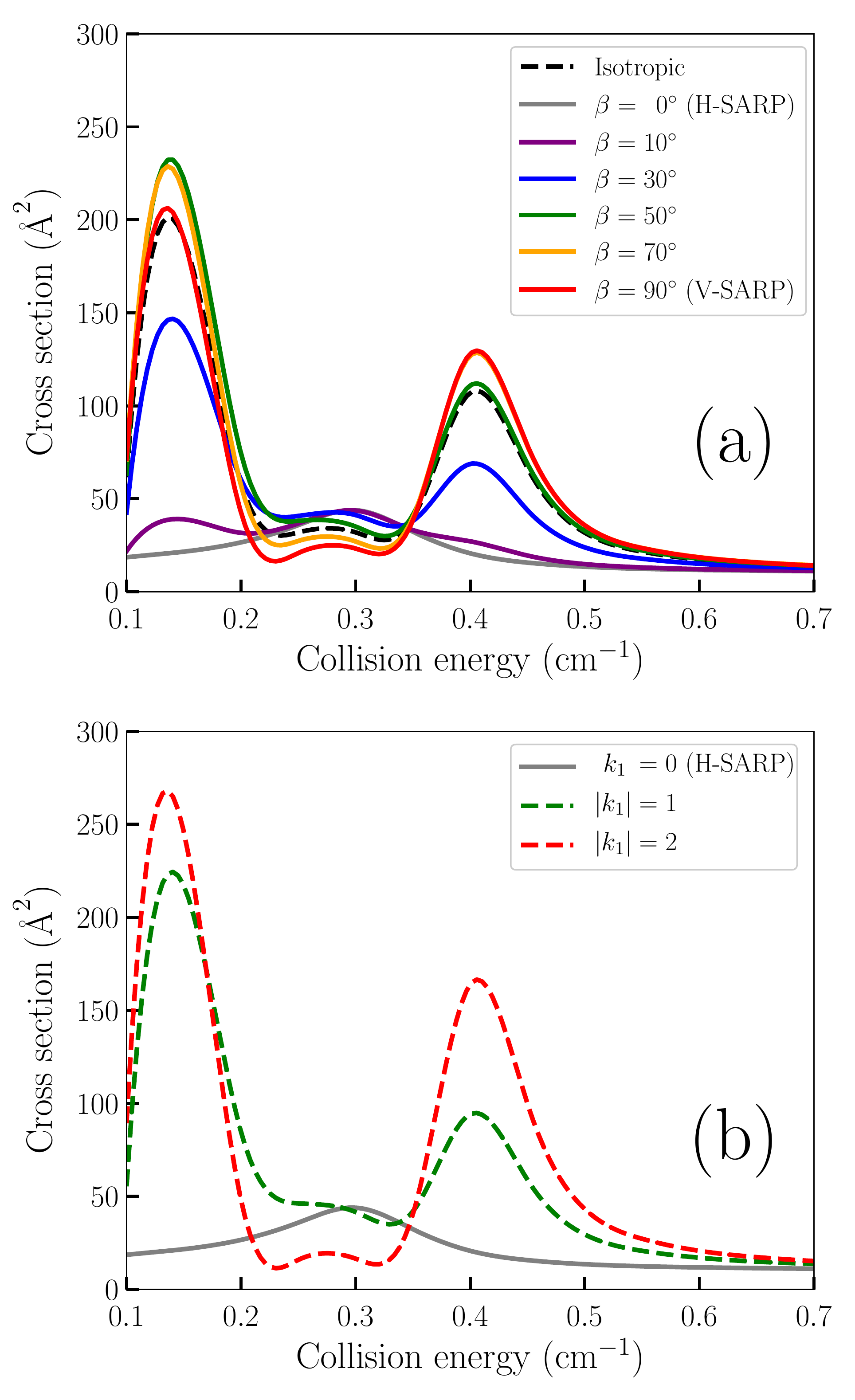}
\end{center}
\caption{
The same as in \Cref{fig_I12F10_steric} but for $\Delta j=-1$ transition.}
\label{fig_I12F11_steric}
\end{figure}

As shown in \cref{fig_I12F11_steric} (a) the angle ($\beta$) between the aligned HCl bond axis and the initial relative velocity vector for the collision has a strong effect on the magnitude of both resonances. However, some notable differences are also visible compared to the $\Delta j_1=-2$ transition (\cref{fig_I12F10_steric} (a)). In \cref{fig_I12F11_steric} (a), the {\it f\,}-wave shape resonance peak increases in intensity with increasing $\beta$ from $0 ^{\circ}$ to $90 ^{\circ}$, just the opposite of the trend observed for the peak in $\Delta j_1=-2$ transition. 
The {\it p\,}-wave peak at lower energy increases in intensity with increasing $\beta$ but it saturates at around $50 ^{\circ}$ before eventually  decreasing with increasing $\beta$ (from $70 ^{\circ}$ to $90 ^{\circ}$). Nevertheless, both resonances exhibit considerable range of control by the alignment preparation. In particular, with small value of $\beta$ ($<$ $10 ^{\circ}$), both peaks are essentially switched off.

To gain more insights into the two resonances \cref{fig_I12F11_steric} (b) presents ICS with different initial HCl orientations. The cross section for $k_1=0$ is the smallest near the resonance peaks and exhibits a small bump near 0.3 cm$^{-1}$, the region in between the two resonances. On the other hand, we observe two intense peaks with $|k_1|=1$ and $2$, revealing the origin of the overlapping resonances. Unlike the $\Delta j_1=-2$ results shown in \cref{fig_I12F10_steric} (b), $|k_1|=2$ gives the largest cross sections while it does not conserve the helicity ($k_1+k_2 \ne k'_1+k'_2$). It means that the resonances for $\Delta j_1=-2$ and $\Delta j_2=-1$ transitions do not share the same mechanism. While the interaction potential does not directly contribute to change in the helicity $k_1+k_2$, the centrifugal term can induce it (Coriolis coupling), and the combined effect of these terms determines the resulting resonance profile. 
Recent studies of molecular reaction and spin-relaxation in the cold energy regime have also pointed out that approximate calculations assuming the conservation of helicity tend to fail in resonance regions \cite{Quemener_restric,Morita_restrict}.


In summary, we have demonstrated 
flexible and robust
control of resonant rotational quenching of HCl in cold collisions with H$_2$  at the level of integral cross sections by preparing HCl in an aligned or oriented state in the $v=1$ vibrational level. 
The most striking result of our study is that even multiple peaks due to resonances associated with disparate partial waves can be controlled, indicating that stereodynamic control is not limited to simple systems with isolated resonances. Our results also reveal the importance of helicity-nonconserving transitions in stereodynamic control of resonant molecular collisions. The analysis based on the initial orientations of HCl rotational state demonstrates the enormous potential controllability in cold molecular collisions, which is tractable with current experimental techniques.

\begin{acknowledgements}
We are grateful to James Croft and Nandini Mukherjee for stimulating discussions. 
This work is supported in part by NSF grant No. PHY-1806334 (N.B.) and ARO MURI grant No. W911NF-19-1-0283 (N.B. and H.G.).
\end{acknowledgements}
\newpage


\bibliography{cite}

\begin{thebibliography}{31}%
\makeatletter
\providecommand \@ifxundefined [1]{%
 \@ifx{#1\undefined}
}%
\providecommand \@ifnum [1]{%
 \ifnum #1\expandafter \@firstoftwo
 \else \expandafter \@secondoftwo
 \fi
}%
\providecommand \@ifx [1]{%
 \ifx #1\expandafter \@firstoftwo
 \else \expandafter \@secondoftwo
 \fi
}%
\providecommand \natexlab [1]{#1}%
\providecommand \enquote  [1]{``#1''}%
\providecommand \bibnamefont  [1]{#1}%
\providecommand \bibfnamefont [1]{#1}%
\providecommand \citenamefont [1]{#1}%
\providecommand \href@noop [0]{\@secondoftwo}%
\providecommand \href [0]{\begingroup \@sanitize@url \@href}%
\providecommand \@href[1]{\@@startlink{#1}\@@href}%
\providecommand \@@href[1]{\endgroup#1\@@endlink}%
\providecommand \@sanitize@url [0]{\catcode `\\12\catcode `\$12\catcode
  `\&12\catcode `\#12\catcode `\^12\catcode `\_12\catcode `\%12\relax}%
\providecommand \@@startlink[1]{}%
\providecommand \@@endlink[0]{}%
\providecommand \url  [0]{\begingroup\@sanitize@url \@url }%
\providecommand \@url [1]{\endgroup\@href {#1}{\urlprefix }}%
\providecommand \urlprefix  [0]{URL }%
\providecommand \Eprint [0]{\href }%
\providecommand \doibase [0]{http://dx.doi.org/}%
\providecommand \selectlanguage [0]{\@gobble}%
\providecommand \bibinfo  [0]{\@secondoftwo}%
\providecommand \bibfield  [0]{\@secondoftwo}%
\providecommand \translation [1]{[#1]}%
\providecommand \BibitemOpen [0]{}%
\providecommand \bibitemStop [0]{}%
\providecommand \bibitemNoStop [0]{.\EOS\space}%
\providecommand \EOS [0]{\spacefactor3000\relax}%
\providecommand \BibitemShut  [1]{\csname bibitem#1\endcsname}%
\let\auto@bib@innerbib\@empty
\bibitem [{\citenamefont {Jones}\ \emph {et~al.}(2006)\citenamefont {Jones},
  \citenamefont {Tiesinga}, \citenamefont {Lett},\ and\ \citenamefont
  {Julienne}}]{Rev_Julienne06}%
  \BibitemOpen
  \bibfield  {author} {\bibinfo {author} {\bibfnamefont {K.~M.}\ \bibnamefont
  {Jones}}, \bibinfo {author} {\bibfnamefont {E.}~\bibnamefont {Tiesinga}},
  \bibinfo {author} {\bibfnamefont {P.~D.}\ \bibnamefont {Lett}}, \ and\
  \bibinfo {author} {\bibfnamefont {P.~S.}\ \bibnamefont {Julienne}},\ }\href
  {\doibase 10.1103/RevModPhys.78.483} {\bibfield  {journal} {\bibinfo
  {journal} {Rev. Mod. Phys.}\ }\textbf {\bibinfo {volume} {78}},\ \bibinfo
  {pages} {483} (\bibinfo {year} {2006})}\BibitemShut {NoStop}%
\bibitem [{\citenamefont {Krems}(2008)}]{Rev_Krems08}%
  \BibitemOpen
  \bibfield  {author} {\bibinfo {author} {\bibfnamefont {R.~V.}\ \bibnamefont
  {Krems}},\ }\href {\doibase 10.1039/B802322K} {\bibfield  {journal} {\bibinfo
   {journal} {Phys. Chem. Chem. Phys.}\ }\textbf {\bibinfo {volume} {10}},\
  \bibinfo {pages} {4079} (\bibinfo {year} {2008})}\BibitemShut {NoStop}%
\bibitem [{\citenamefont {Carr}\ \emph {et~al.}(2009)\citenamefont {Carr},
  \citenamefont {DeMille}, \citenamefont {Krems},\ and\ \citenamefont
  {Ye}}]{Rev_Carr08}%
  \BibitemOpen
  \bibfield  {author} {\bibinfo {author} {\bibfnamefont {L.~D.}\ \bibnamefont
  {Carr}}, \bibinfo {author} {\bibfnamefont {D.}~\bibnamefont {DeMille}},
  \bibinfo {author} {\bibfnamefont {R.~V.}\ \bibnamefont {Krems}}, \ and\
  \bibinfo {author} {\bibfnamefont {J.}~\bibnamefont {Ye}},\ }\href
  {http://stacks.iop.org/1367-2630/11/i=5/a=055049} {\bibfield  {journal}
  {\bibinfo  {journal} {New J. Phys.}\ }\textbf {\bibinfo {volume} {11}},\
  \bibinfo {pages} {055049} (\bibinfo {year} {2009})}\BibitemShut {NoStop}%
\bibitem [{\citenamefont {Naulin}\ and\ \citenamefont
  {Costes}(2014)}]{Rev_Costes_2014}%
  \BibitemOpen
  \bibfield  {author} {\bibinfo {author} {\bibfnamefont {C.}~\bibnamefont
  {Naulin}}\ and\ \bibinfo {author} {\bibfnamefont {M.}~\bibnamefont
  {Costes}},\ }\href {\doibase 10.1080/0144235X.2014.957565} {\bibfield
  {journal} {\bibinfo  {journal} {Int. Rev. Phys. Chem.}\ }\textbf {\bibinfo
  {volume} {33}},\ \bibinfo {pages} {427} (\bibinfo {year} {2014})}\BibitemShut
  {NoStop}%
\bibitem [{\citenamefont {Stuhl}\ \emph {et~al.}(2014)\citenamefont {Stuhl},
  \citenamefont {Hummon},\ and\ \citenamefont {Ye}}]{Rev_Ye2014}%
  \BibitemOpen
  \bibfield  {author} {\bibinfo {author} {\bibfnamefont {B.~K.}\ \bibnamefont
  {Stuhl}}, \bibinfo {author} {\bibfnamefont {M.~T.}\ \bibnamefont {Hummon}}, \
  and\ \bibinfo {author} {\bibfnamefont {J.}~\bibnamefont {Ye}},\ }\href
  {\doibase 10.1146/annurev-physchem-040513-103744} {\bibfield  {journal}
  {\bibinfo  {journal} {Ann. Rev. Phys. Chem.}\ }\textbf {\bibinfo {volume}
  {65}},\ \bibinfo {pages} {501} (\bibinfo {year} {2014})}\BibitemShut
  {NoStop}%
\bibitem [{\citenamefont {Balakrishnan}(2016)}]{Rev_Bala16}%
  \BibitemOpen
  \bibfield  {author} {\bibinfo {author} {\bibfnamefont {N.}~\bibnamefont
  {Balakrishnan}},\ }\href {\doibase 10.1063/1.4964096} {\bibfield  {journal}
  {\bibinfo  {journal} {J. Chem. Phys.}\ }\textbf {\bibinfo {volume} {145}},\
  \bibinfo {pages} {150901} (\bibinfo {year} {2016})}\BibitemShut {NoStop}%
\bibitem [{\citenamefont {Bohn}\ \emph {et~al.}(2017)\citenamefont {Bohn},
  \citenamefont {Rey},\ and\ \citenamefont {Ye}}]{Rev_Bohn17}%
  \BibitemOpen
  \bibfield  {author} {\bibinfo {author} {\bibfnamefont {J.~L.}\ \bibnamefont
  {Bohn}}, \bibinfo {author} {\bibfnamefont {A.~M.}\ \bibnamefont {Rey}}, \
  and\ \bibinfo {author} {\bibfnamefont {J.}~\bibnamefont {Ye}},\ }\href
  {\doibase 10.1126/science.aam6299} {\bibfield  {journal} {\bibinfo  {journal}
  {Science}\ }\textbf {\bibinfo {volume} {357}},\ \bibinfo {pages} {1002}
  (\bibinfo {year} {2017})}\BibitemShut {NoStop}%
\bibitem [{\citenamefont {Safronova}\ \emph {et~al.}(2018)\citenamefont
  {Safronova}, \citenamefont {Budker}, \citenamefont {DeMille}, \citenamefont
  {Kimball}, \citenamefont {Derevianko},\ and\ \citenamefont
  {Clark}}]{Rev_Safronova18}%
  \BibitemOpen
  \bibfield  {author} {\bibinfo {author} {\bibfnamefont {M.~S.}\ \bibnamefont
  {Safronova}}, \bibinfo {author} {\bibfnamefont {D.}~\bibnamefont {Budker}},
  \bibinfo {author} {\bibfnamefont {D.}~\bibnamefont {DeMille}}, \bibinfo
  {author} {\bibfnamefont {D.~F.~J.}\ \bibnamefont {Kimball}}, \bibinfo
  {author} {\bibfnamefont {A.}~\bibnamefont {Derevianko}}, \ and\ \bibinfo
  {author} {\bibfnamefont {C.~W.}\ \bibnamefont {Clark}},\ }\href {\doibase
  10.1103/RevModPhys.90.025008} {\bibfield  {journal} {\bibinfo  {journal}
  {Rev. Mod. Phys.}\ }\textbf {\bibinfo {volume} {90}},\ \bibinfo {pages}
  {025008} (\bibinfo {year} {2018})}\BibitemShut {NoStop}%
\bibitem [{\citenamefont {Shagam}\ \emph {et~al.}(2015)\citenamefont {Shagam},
  \citenamefont {Klein}, \citenamefont {Skomorowski}, \citenamefont {Yun},
  \citenamefont {Averbukh}, \citenamefont {Koch},\ and\ \citenamefont
  {Narevicius}}]{Merge}%
  \BibitemOpen
  \bibfield  {author} {\bibinfo {author} {\bibfnamefont {Y.}~\bibnamefont
  {Shagam}}, \bibinfo {author} {\bibfnamefont {A.}~\bibnamefont {Klein}},
  \bibinfo {author} {\bibfnamefont {W.}~\bibnamefont {Skomorowski}}, \bibinfo
  {author} {\bibfnamefont {R.}~\bibnamefont {Yun}}, \bibinfo {author}
  {\bibfnamefont {V.}~\bibnamefont {Averbukh}}, \bibinfo {author}
  {\bibfnamefont {C.~P.}\ \bibnamefont {Koch}}, \ and\ \bibinfo {author}
  {\bibfnamefont {E.}~\bibnamefont {Narevicius}},\ }\href {\doibase
  10.1038/nchem.2359} {\bibfield  {journal} {\bibinfo  {journal} {Nat. Chem.}\
  }\textbf {\bibinfo {volume} {7}},\ \bibinfo {pages} {921} (\bibinfo {year}
  {2015})}\BibitemShut {NoStop}%
\bibitem [{\citenamefont {Segev}\ \emph {et~al.}(2019)\citenamefont {Segev},
  \citenamefont {Pitzer}, \citenamefont {Karpov}, \citenamefont {Akerman},
  \citenamefont {Narevicius},\ and\ \citenamefont {Narevicius}}]{Superconduct}%
  \BibitemOpen
  \bibfield  {author} {\bibinfo {author} {\bibfnamefont {Y.}~\bibnamefont
  {Segev}}, \bibinfo {author} {\bibfnamefont {M.}~\bibnamefont {Pitzer}},
  \bibinfo {author} {\bibfnamefont {M.}~\bibnamefont {Karpov}}, \bibinfo
  {author} {\bibfnamefont {N.}~\bibnamefont {Akerman}}, \bibinfo {author}
  {\bibfnamefont {J.}~\bibnamefont {Narevicius}}, \ and\ \bibinfo {author}
  {\bibfnamefont {E.}~\bibnamefont {Narevicius}},\ }\href {\doibase
  10.1038/s41586-019-1446-2} {\bibfield  {journal} {\bibinfo  {journal}
  {Nature}\ }\textbf {\bibinfo {volume} {572}},\ \bibinfo {pages} {189}
  (\bibinfo {year} {2019})}\BibitemShut {NoStop}%
\bibitem [{\citenamefont {Perreault}\ \emph {et~al.}(2017)\citenamefont
  {Perreault}, \citenamefont {Mukherjee},\ and\ \citenamefont
  {Zare}}]{2017_Science_Perreault}%
  \BibitemOpen
  \bibfield  {author} {\bibinfo {author} {\bibfnamefont {W.~E.}\ \bibnamefont
  {Perreault}}, \bibinfo {author} {\bibfnamefont {N.}~\bibnamefont
  {Mukherjee}}, \ and\ \bibinfo {author} {\bibfnamefont {R.~N.}\ \bibnamefont
  {Zare}},\ }\href {\doibase 10.1126/science.aao3116} {\bibfield  {journal}
  {\bibinfo  {journal} {Science}\ }\textbf {\bibinfo {volume} {358}},\ \bibinfo
  {pages} {356} (\bibinfo {year} {2017})}\BibitemShut {NoStop}%
\bibitem [{\citenamefont {Perreault}\ \emph {et~al.}(2018)\citenamefont
  {Perreault}, \citenamefont {Mukherjee},\ and\ \citenamefont
  {Zare}}]{2018_NatChem_Perreault}%
  \BibitemOpen
  \bibfield  {author} {\bibinfo {author} {\bibfnamefont {W.~E.}\ \bibnamefont
  {Perreault}}, \bibinfo {author} {\bibfnamefont {N.}~\bibnamefont
  {Mukherjee}}, \ and\ \bibinfo {author} {\bibfnamefont {R.~N.}\ \bibnamefont
  {Zare}},\ }\href {\doibase 10.1038/s41557-018-0028-5} {\bibfield  {journal}
  {\bibinfo  {journal} {Nat. Chem.}\ }\textbf {\bibinfo {volume} {10}},\
  \bibinfo {pages} {561} (\bibinfo {year} {2018})}\BibitemShut {NoStop}%
\bibitem [{\citenamefont {Perreault}\ \emph
  {et~al.}(2019{\natexlab{a}})\citenamefont {Perreault}, \citenamefont
  {Mukherjee},\ and\ \citenamefont {Zare}}]{SARP_HD-He}%
  \BibitemOpen
  \bibfield  {author} {\bibinfo {author} {\bibfnamefont {W.~E.}\ \bibnamefont
  {Perreault}}, \bibinfo {author} {\bibfnamefont {N.}~\bibnamefont
  {Mukherjee}}, \ and\ \bibinfo {author} {\bibfnamefont {R.~N.}\ \bibnamefont
  {Zare}},\ }\href {\doibase 10.1063/1.5096531} {\bibfield  {journal} {\bibinfo
   {journal} {J. Chem. Phys.}\ }\textbf {\bibinfo {volume} {150}},\ \bibinfo
  {pages} {174301} (\bibinfo {year} {2019}{\natexlab{a}})}\BibitemShut
  {NoStop}%
\bibitem [{\citenamefont {Zare}(1998)}]{Zare_Rev}%
  \BibitemOpen
  \bibfield  {author} {\bibinfo {author} {\bibfnamefont {R.~N.}\ \bibnamefont
  {Zare}},\ }\href {\doibase 10.1126/science.279.5358.1875} {\bibfield
  {journal} {\bibinfo  {journal} {Science}\ }\textbf {\bibinfo {volume}
  {279}},\ \bibinfo {pages} {1875} (\bibinfo {year} {1998})}\BibitemShut
  {NoStop}%
\bibitem [{\citenamefont {Orr-Ewing}(1996)}]{Orr-Ewing}%
  \BibitemOpen
  \bibfield  {author} {\bibinfo {author} {\bibfnamefont {A.~J.}\ \bibnamefont
  {Orr-Ewing}},\ }\href {\doibase 10.1039/FT9969200881} {\bibfield  {journal}
  {\bibinfo  {journal} {J. Chem. Soc.{,} Faraday Trans.}\ }\textbf {\bibinfo
  {volume} {92}},\ \bibinfo {pages} {881} (\bibinfo {year} {1996})}\BibitemShut
  {NoStop}%
\bibitem [{\citenamefont {Aldegunde}\ \emph {et~al.}(2005)\citenamefont
  {Aldegunde}, \citenamefont {de~Miranda}, \citenamefont {Haigh}, \citenamefont
  {Kendrick}, \citenamefont {S{\'a}ez-R{\'a}banos},\ and\ \citenamefont
  {Aoiz}}]{Aldegunde}%
  \BibitemOpen
  \bibfield  {author} {\bibinfo {author} {\bibfnamefont {J.}~\bibnamefont
  {Aldegunde}}, \bibinfo {author} {\bibfnamefont {M.~P.}\ \bibnamefont
  {de~Miranda}}, \bibinfo {author} {\bibfnamefont {J.~M.}\ \bibnamefont
  {Haigh}}, \bibinfo {author} {\bibfnamefont {B.~K.}\ \bibnamefont {Kendrick}},
  \bibinfo {author} {\bibfnamefont {V.}~\bibnamefont {S{\'a}ez-R{\'a}banos}}, \
  and\ \bibinfo {author} {\bibfnamefont {F.~J.}\ \bibnamefont {Aoiz}},\ }\href
  {\doibase 10.1021/jp0512208} {\bibfield  {journal} {\bibinfo  {journal} {J.
  Phys. Chem. A}\ }\textbf {\bibinfo {volume} {109}},\ \bibinfo {pages} {6200}
  (\bibinfo {year} {2005})}\BibitemShut {NoStop}%
\bibitem [{\citenamefont {Wang}\ \emph {et~al.}(2012)\citenamefont {Wang},
  \citenamefont {Liu},\ and\ \citenamefont {Rakitzis}}]{Kopin}%
  \BibitemOpen
  \bibfield  {author} {\bibinfo {author} {\bibfnamefont {F.}~\bibnamefont
  {Wang}}, \bibinfo {author} {\bibfnamefont {K.}~\bibnamefont {Liu}}, \ and\
  \bibinfo {author} {\bibfnamefont {T.~P.}\ \bibnamefont {Rakitzis}},\ }\href
  {\doibase 10.1038/nchem.1383} {\bibfield  {journal} {\bibinfo  {journal}
  {Nat. Chem.}\ }\textbf {\bibinfo {volume} {4}},\ \bibinfo {pages} {636}
  (\bibinfo {year} {2012})}\BibitemShut {NoStop}%
\bibitem [{\citenamefont {Dong}\ \emph
  {et~al.}(2013{\natexlab{a}})\citenamefont {Dong}, \citenamefont {Mukherjee},\
  and\ \citenamefont {Zare}}]{SARP_1}%
  \BibitemOpen
  \bibfield  {author} {\bibinfo {author} {\bibfnamefont {W.}~\bibnamefont
  {Dong}}, \bibinfo {author} {\bibfnamefont {N.}~\bibnamefont {Mukherjee}}, \
  and\ \bibinfo {author} {\bibfnamefont {R.~N.}\ \bibnamefont {Zare}},\ }\href
  {\doibase 10.1063/1.4818526} {\bibfield  {journal} {\bibinfo  {journal} {J.
  Chem. Phys.}\ }\textbf {\bibinfo {volume} {139}},\ \bibinfo {pages} {074204}
  (\bibinfo {year} {2013}{\natexlab{a}})}\BibitemShut {NoStop}%
\bibitem [{\citenamefont {Mukherjee}\ \emph {et~al.}(2014)\citenamefont
  {Mukherjee}, \citenamefont {Dong},\ and\ \citenamefont {Zare}}]{SARP_2}%
  \BibitemOpen
  \bibfield  {author} {\bibinfo {author} {\bibfnamefont {N.}~\bibnamefont
  {Mukherjee}}, \bibinfo {author} {\bibfnamefont {W.}~\bibnamefont {Dong}}, \
  and\ \bibinfo {author} {\bibfnamefont {R.~N.}\ \bibnamefont {Zare}},\ }\href
  {\doibase 10.1063/1.4865131} {\bibfield  {journal} {\bibinfo  {journal} {J.
  Chem. Phys.}\ }\textbf {\bibinfo {volume} {140}},\ \bibinfo {pages} {074201}
  (\bibinfo {year} {2014})}\BibitemShut {NoStop}%
\bibitem [{\citenamefont {Croft}\ \emph {et~al.}(2018)\citenamefont {Croft},
  \citenamefont {Balakrishnan}, \citenamefont {Huang},\ and\ \citenamefont
  {Guo}}]{2018_PRL_Croft}%
  \BibitemOpen
  \bibfield  {author} {\bibinfo {author} {\bibfnamefont {J.~F.~E.}\
  \bibnamefont {Croft}}, \bibinfo {author} {\bibfnamefont {N.}~\bibnamefont
  {Balakrishnan}}, \bibinfo {author} {\bibfnamefont {M.}~\bibnamefont {Huang}},
  \ and\ \bibinfo {author} {\bibfnamefont {H.}~\bibnamefont {Guo}},\ }\href
  {\doibase 10.1103/PhysRevLett.121.113401} {\bibfield  {journal} {\bibinfo
  {journal} {Phys. Rev. Lett.}\ }\textbf {\bibinfo {volume} {121}},\ \bibinfo
  {pages} {113401} (\bibinfo {year} {2018})}\BibitemShut {NoStop}%
\bibitem [{\citenamefont {Croft}\ and\ \citenamefont
  {Balakrishnan}(2019)}]{2019_JCP_Croft}%
  \BibitemOpen
  \bibfield  {author} {\bibinfo {author} {\bibfnamefont {J.~F.~E.}\
  \bibnamefont {Croft}}\ and\ \bibinfo {author} {\bibfnamefont
  {N.}~\bibnamefont {Balakrishnan}},\ }\href {\doibase 10.1063/1.5091576}
  {\bibfield  {journal} {\bibinfo  {journal} {J. Chem. Phys.}\ }\textbf
  {\bibinfo {volume} {150}},\ \bibinfo {pages} {164302} (\bibinfo {year}
  {2019})}\BibitemShut {NoStop}%
\bibitem [{\citenamefont {Jambrina}\ \emph {et~al.}(2019)\citenamefont
  {Jambrina}, \citenamefont {Croft}, \citenamefont {Guo}, \citenamefont
  {Brouard}, \citenamefont {Balakrishnan},\ and\ \citenamefont
  {Aoiz}}]{2019_PRL_Jambrina}%
  \BibitemOpen
  \bibfield  {author} {\bibinfo {author} {\bibfnamefont {P.~G.}\ \bibnamefont
  {Jambrina}}, \bibinfo {author} {\bibfnamefont {J.~F.~E.}\ \bibnamefont
  {Croft}}, \bibinfo {author} {\bibfnamefont {H.}~\bibnamefont {Guo}}, \bibinfo
  {author} {\bibfnamefont {M.}~\bibnamefont {Brouard}}, \bibinfo {author}
  {\bibfnamefont {N.}~\bibnamefont {Balakrishnan}}, \ and\ \bibinfo {author}
  {\bibfnamefont {F.~J.}\ \bibnamefont {Aoiz}},\ }\href {\doibase
  10.1103/PhysRevLett.123.043401} {\bibfield  {journal} {\bibinfo  {journal}
  {Phys. Rev. Lett.}\ }\textbf {\bibinfo {volume} {123}},\ \bibinfo {pages}
  {043401} (\bibinfo {year} {2019})}\BibitemShut {NoStop}%
\bibitem [{\citenamefont {Dong}\ \emph
  {et~al.}(2013{\natexlab{b}})\citenamefont {Dong}, \citenamefont {Mukherjee},\
  and\ \citenamefont {Zare}}]{SARP_H2}%
  \BibitemOpen
  \bibfield  {author} {\bibinfo {author} {\bibfnamefont {W.}~\bibnamefont
  {Dong}}, \bibinfo {author} {\bibfnamefont {N.}~\bibnamefont {Mukherjee}}, \
  and\ \bibinfo {author} {\bibfnamefont {R.~N.}\ \bibnamefont {Zare}},\ }\href
  {\doibase 10.1063/1.4818526} {\bibfield  {journal} {\bibinfo  {journal} {J.
  Chem Phys.}\ }\textbf {\bibinfo {volume} {139}},\ \bibinfo {pages} {074204}
  (\bibinfo {year} {2013}{\natexlab{b}})}\BibitemShut {NoStop}%
\bibitem [{\citenamefont {Perreault}\ \emph
  {et~al.}(2019{\natexlab{b}})\citenamefont {Perreault}, \citenamefont
  {Mukherjee},\ and\ \citenamefont {Zare}}]{SARP_D2}%
  \BibitemOpen
  \bibfield  {author} {\bibinfo {author} {\bibfnamefont {W.~E.}\ \bibnamefont
  {Perreault}}, \bibinfo {author} {\bibfnamefont {N.}~\bibnamefont
  {Mukherjee}}, \ and\ \bibinfo {author} {\bibfnamefont {R.~N.}\ \bibnamefont
  {Zare}},\ }\href {\doibase 10.1063/1.5109261} {\bibfield  {journal} {\bibinfo
   {journal} {J. Chem. Phys.}\ }\textbf {\bibinfo {volume} {150}},\ \bibinfo
  {pages} {234201} (\bibinfo {year} {2019}{\natexlab{b}})}\BibitemShut
  {NoStop}%
\bibitem [{\citenamefont {Yao}\ \emph {et~al.}(2019)\citenamefont {Yao},
  \citenamefont {Morita}, \citenamefont {Xie}, \citenamefont {Balakrishnan},\
  and\ \citenamefont {Guo}}]{H2HCl_PES}%
  \BibitemOpen
  \bibfield  {author} {\bibinfo {author} {\bibfnamefont {Q.}~\bibnamefont
  {Yao}}, \bibinfo {author} {\bibfnamefont {M.}~\bibnamefont {Morita}},
  \bibinfo {author} {\bibfnamefont {C.}~\bibnamefont {Xie}}, \bibinfo {author}
  {\bibfnamefont {N.}~\bibnamefont {Balakrishnan}}, \ and\ \bibinfo {author}
  {\bibfnamefont {H.}~\bibnamefont {Guo}},\ }\href {\doibase
  10.1021/acs.jpca.9b05958} {\bibfield  {journal} {\bibinfo  {journal} {J.
  Phys. Chem. A}\ }\textbf {\bibinfo {volume} {123}},\ \bibinfo {pages} {6578}
  (\bibinfo {year} {2019})}\BibitemShut {NoStop}%
\bibitem [{\citenamefont {Krems}(2006)}]{TwoBC}%
  \BibitemOpen
  \bibfield  {author} {\bibinfo {author} {\bibfnamefont {R.~V.}\ \bibnamefont
  {Krems}},\ }\href@noop {} {\emph {\bibinfo {title} {TwoBC Quantum Scattering
  Program}}},\ \bibinfo {organization} {Vancouver: University of British
  Columbia} (\bibinfo {year} {2006})\BibitemShut {NoStop}%
\bibitem [{\citenamefont {Quéméner}\ and\ \citenamefont
  {Balakrishnan}(2009)}]{CC_3}%
  \BibitemOpen
  \bibfield  {author} {\bibinfo {author} {\bibfnamefont {G.}~\bibnamefont
  {Quéméner}}\ and\ \bibinfo {author} {\bibfnamefont {N.}~\bibnamefont
  {Balakrishnan}},\ }\href {\doibase 10.1063/1.3081225} {\bibfield  {journal}
  {\bibinfo  {journal} {J. Chem. Phys.}\ }\textbf {\bibinfo {volume} {130}},\
  \bibinfo {pages} {114303} (\bibinfo {year} {2009})}\BibitemShut {NoStop}%
\bibitem [{\citenamefont {Schaefer}\ and\ \citenamefont {Meyer}(1979)}]{DCS}%
  \BibitemOpen
  \bibfield  {author} {\bibinfo {author} {\bibfnamefont {J.}~\bibnamefont
  {Schaefer}}\ and\ \bibinfo {author} {\bibfnamefont {W.}~\bibnamefont
  {Meyer}},\ }\href {\doibase 10.1063/1.437196} {\bibfield  {journal} {\bibinfo
   {journal} {J. Chem. Phys.}\ }\textbf {\bibinfo {volume} {70}},\ \bibinfo
  {pages} {344} (\bibinfo {year} {1979})}\BibitemShut {NoStop}%
\bibitem [{SM()}]{SM}%
  \BibitemOpen
  \href@noop {} {}\bibinfo {note} {See Supplemental Material at [
  http://link.***.org/ supplemental/XXX/ ] for squared Wigner's reduced
  rotational matrix elements, adiabatic and diabatic potential energy curves,
  differential cross section, and integral cross section for minor transitions
  associated with vibrational relaxation of HCl.}\BibitemShut {Stop}%
\bibitem [{\citenamefont {Quéméner}\ and\ \citenamefont
  {Balakrishnan}(2008)}]{Quemener_restric}%
  \BibitemOpen
  \bibfield  {author} {\bibinfo {author} {\bibfnamefont {G.}~\bibnamefont
  {Quéméner}}\ and\ \bibinfo {author} {\bibfnamefont {N.}~\bibnamefont
  {Balakrishnan}},\ }\href {\doibase 10.1063/1.2928804} {\bibfield  {journal}
  {\bibinfo  {journal} {J. Chem. Phys.}\ }\textbf {\bibinfo {volume} {128}},\
  \bibinfo {pages} {224304} (\bibinfo {year} {2008})}\BibitemShut {NoStop}%
\bibitem [{\citenamefont {Morita}\ and\ \citenamefont
  {Tscherbul}(2019)}]{Morita_restrict}%
  \BibitemOpen
  \bibfield  {author} {\bibinfo {author} {\bibfnamefont {M.}~\bibnamefont
  {Morita}}\ and\ \bibinfo {author} {\bibfnamefont {T.~V.}\ \bibnamefont
  {Tscherbul}},\ }\href {\doibase 10.1063/1.5047063} {\bibfield  {journal}
  {\bibinfo  {journal} {J. Chem. Phys.}\ }\textbf {\bibinfo {volume} {150}},\
  \bibinfo {pages} {074110} (\bibinfo {year} {2019})}\BibitemShut {NoStop}%
\end{thebibliography}%



\clearpage
\onecolumngrid
\vspace{\columnsep}

\newcolumntype{Y}{>{\centering\arraybackslash}X}
\newcolumntype{Z}{>{\raggedleft\arraybackslash}X}

\setcounter{figure}{0}
\setcounter{equation}{0}
\setcounter{page}{1}

\renewcommand{\thepage}{S\arabic{page}}
\renewcommand{\thefigure}{S\arabic{figure}}
\renewcommand{\theequation}{S\arabic{equation}}

\onecolumngrid

\begin{center}
	\textbf{\huge Supplemental Material}
\end{center}

\begin{center}
\text{ \Large Stereodynamic control of overlapping resonances with aligned cold molecules} 
\end{center}

\begin{center}
\text{ \large Masato Morita, Qian Yao, Changjian Xie, Hua Guo and Naduvalath Balakrishnan} 
\end{center}
\begin{center}
\end{center}


\begin{center}
	\textbf{\large S-I.\ Weighting factor for the aligned HCl}
\end{center}

\begin{figure}[h!]
\begin{center}
\includegraphics[scale=0.45
]{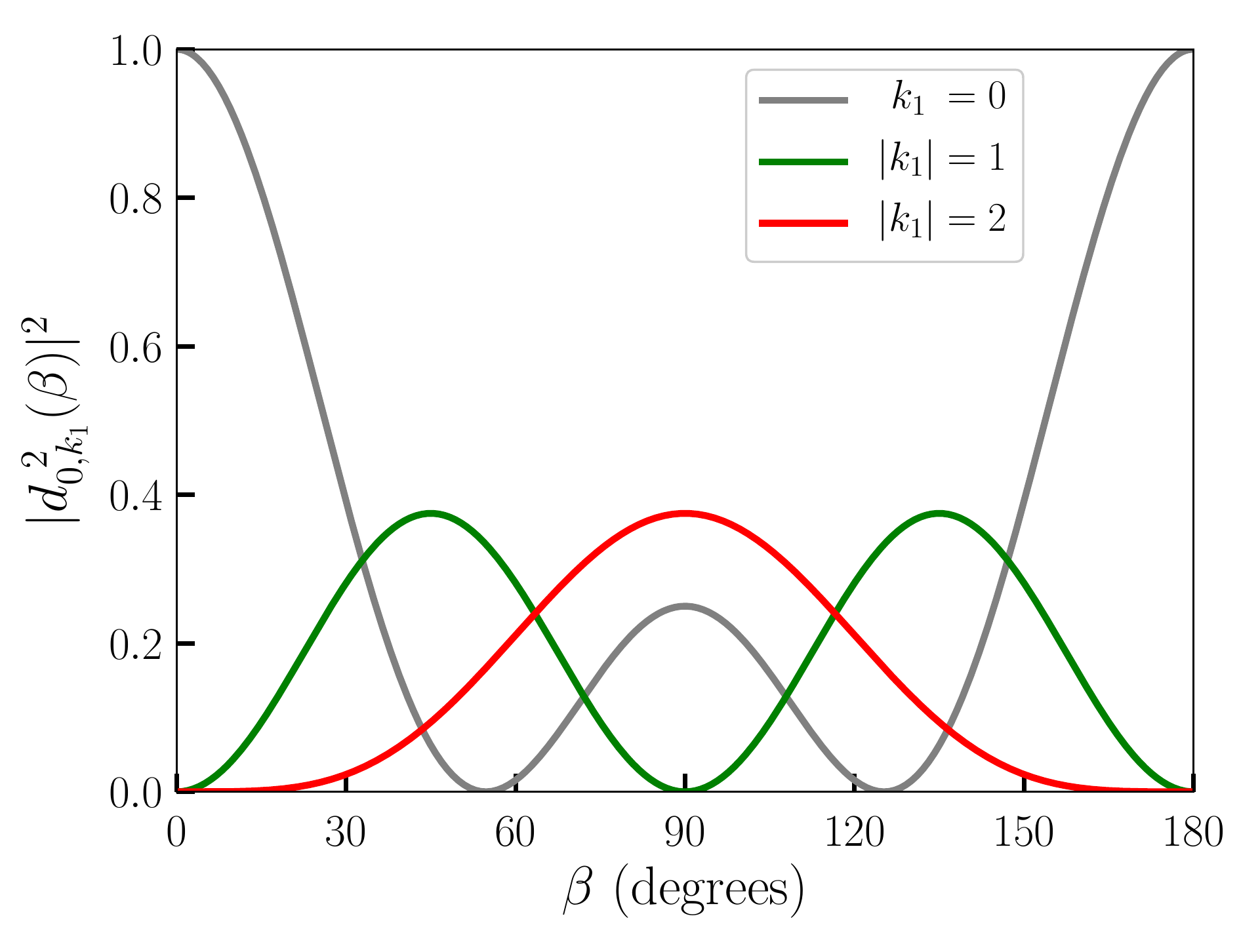}
\end{center}
\caption{Weighting factor $|d^{2}_{0,k_1}(\beta)|^2$ for the aligned HCl ($j_1=2$, 
$\tilde{m}_1=0$) as a function of $\beta$. The quantum number $k_1$ is the projection of $j_1$ onto the initial relative velocity vector between the collision partners. The angle between the aligned HCl molecular bond axis and the initial relative velocity vector is denoted by $\beta$ .\\ 
}
\label{fig_d}
\end{figure}

As outlined in the main text, the HCl bond axis is preferentially aligned parallel to the linear polarization direction of the SARP laser (pump and Stokes laser pulses which intersect the laboratory fixed molecular beam) and the rotational state is written as 
$|j_1,\tilde{m}_1=0>$, where $\tilde{m}_1$ 
is the projection of $j_1$ onto an axis parallel to the polarization of the SARP laser. The state can be expressed as a coherent superposition of rotational states $|j_1,k_1> (k_1=-j_1, -j_1+1,...,j_1+1, j_1)$, where $k_1$ is the projection of $j_1$ onto 
the z-axis of the body-fixed coordinate frame which is parallel to the relative velocity vector for the collision. 
As a function of the angle ($\beta$) between the (preferentially aligned) HCl molecular bond axis and the initial relative velocity vector for the collision, the coefficients for the superposition are given by Wigner's reduced rotation matrix as $d^{j_1}_{0,k_1}(\beta)$. The differential cross section as a function of $\theta$ is expressed using the squared value $|d^{j_1}_{0,k_1}(\beta)|^2$ as the weighting factor for each $|q|^2$ term associated with $k_1$ as long as the azimuthal angle ($\phi$) dependence of the scattering is averaged over by taking the integral as mentioned in the main text. 
Since the cross section is given as the sum of positive valued terms, the magnitude of the weighting factor is essential to analyze the the cross section. For $j_1=2$, the weighting factor $|d^{2}_{0,k_1}(\beta)|^2$ associated with 5 possible $k_1$ terms ($k_1=-2,-1,0,1,2$) is shown below (\cref{fig_d}). An essentially identical figure is shown in Ref.~\cite{2019_JCP_Croft} for the study of HD+H$_2$. For example,  $\beta=0 ^{\circ}$ corresponds to  H-SARP, and the term associated with $k_1=0$ has a weighting factor of 1.0 because the HCl molecular bond axis is preferentially aligned parallel to the initial relative velocity vector for the collision. For $\beta=90 ^{\circ}$ corresponding to the V-SARP preparation, the terms associated with $k_1=\pm1$ make no contribution.  

\begin{center}
	\textbf{\large S-I\hspace{-.1em}I.\ Adiabatic and diabatic potential energy curves}
\end{center}

The collision dynamics of HCl+H$_2$ is more complex  compared with previously studied HD+H$_2$ \cite{2018_PRL_Croft} due to the larger anisotropy and deeper well depth of the HCl+H$_2$ PES and the smaller rotational constant of HCl. These characteristics give rise to strong interchannel coupling and high density of states/resonances. 

Quantum scattering calculations are performed to obtain the scattering $S$-matrix using the  close-coupling (CC) method on a full six-dimensional (6D) {\it ab initio} PES obtained in our recent work \cite{H2HCl_PES}. An analytical fit of the PES using  the permutation invariant polynomial neural network (PIP-NN) method was performed based on the grid points calculated with CCSD(T)-F12b/aug-cc-pVQZ (AVQZ). 
The CC equations for the HCl+H$_2$ collision complex  obtained using the total angular momentum basis set $|JM(lj_{12}(j_1j_2))>$ for the angular variables in a space-fixed (SF) coordinate frame are numerically solved by the log-derivative propagation method using the TwoBC code. 

Here, we show the adiabatic and diabatic potential energy curves for HCl+H$_2$ and provide further insights into the strong coupling at short-range during the collision. 
The diabatic potential energy curves with the basis set \begin{equation}\label{eq:basis}
    r_1^{-1}\phi^{j_1}_{v_1}(r_1)    r_2^{-1}\chi^{j_2}_{v_2}(r_2) |JM(lj_{12}(j_1j_2))>,
\end{equation} where $\phi^{j_1}_{v_1}$ and $\chi^{j_2}_{v_2}$ are the radial part of the eigenfunctions of rovibrational states of HCl and H$_2$, are defined as the diagonal elements of the effective potential $W(R)$ in \cref{eq:Veff} below. On the other hand, the adiabatic potential energy curves are given as the eigenvalues obtained by diagonalizing the $W(R)$ matrix at each value of the parameter $R$, thus, in principle, the adiabatic potential energy curves are independent of basis sets. The matrix element in $W(R)$ in the basis set (\cref{eq:basis}) is written as 
\begin{equation}\label{eq:Veff}
\begin{split}
    W^{J,\epsilon_I}_{\alpha j_{12} l, \alpha' j'_{12} l'}(R) 
    & = \frac{l(l+1)\hbar^2}{2 \mu R^2} \delta_{\alpha j_{12} l,\alpha' j'_{12} l'} 
     + V^{J,\epsilon_I}_{\alpha j_{12} l, \alpha' j'_{12} l'}(R) \\
    & = [E^0_{\alpha}+\frac{l(l+1)\hbar^2}{2 \mu R^2} ] \delta_{\alpha j_{12} l,\alpha' j'_{12} l'}  
     + {U_\text{int}}^{J,\epsilon_I}_{\alpha j_{12} l, \alpha' j'_{12} l'}(R),  
\end{split}
\end{equation}
where $\alpha\equiv v_1j_1v_2j_2$ refers to the combined molecular state, $\epsilon_I$ is the inversion parity defined as $(-1)^{l+j_1+j_2}$, $\mu$ is the reduced mass for the molecular collision, $V(R)$ is the potential energy of the tetra-atomic system, $E^0_\alpha$ denote the channel energies for the separated molecules whose quantum states are specified by $\alpha$ and $U_\text{int}(R)$ is the interaction potential between HCl and H$_2$. Here, we shows the adiabatic and diabatic potential energy curves obtained with $J=3$ and $\epsilon_I=-1$ because the incoming {\it p\,}-wave ($l=1$), which generates the shape resonance discussed in the main text, is included for the initial state $\alpha\equiv(1,2,0,0)$ in this study.

\begin{figure}[htbp]
  \begin{center}
    \begin{tabular}{c}

      \begin{minipage}{0.5\hsize}
        \begin{center}
          \includegraphics[ width=9.0cm]{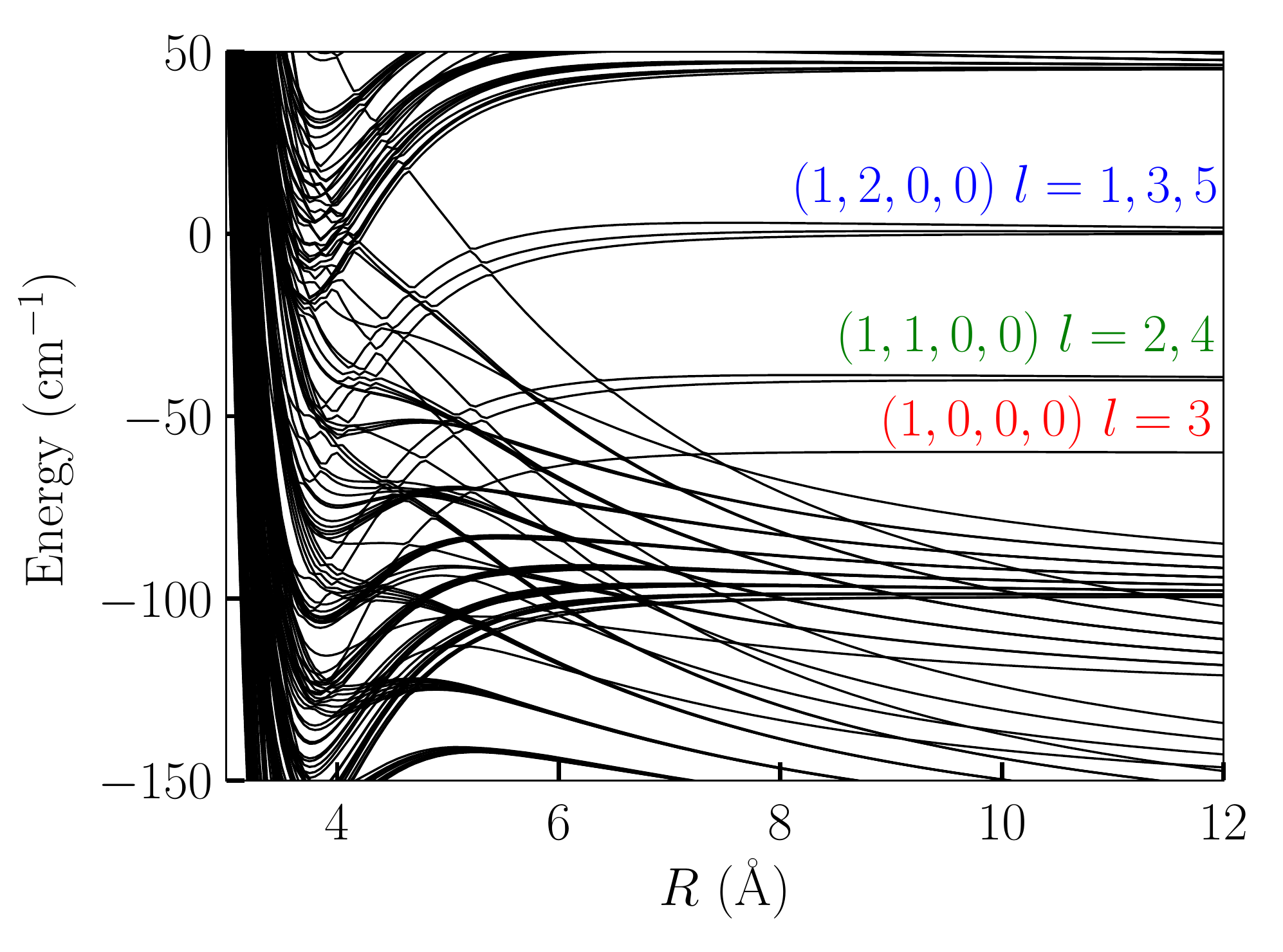}
        \end{center}
      \end{minipage}

      \begin{minipage}{0.5\hsize}
        \begin{center}
          \includegraphics[ width=9.0cm]{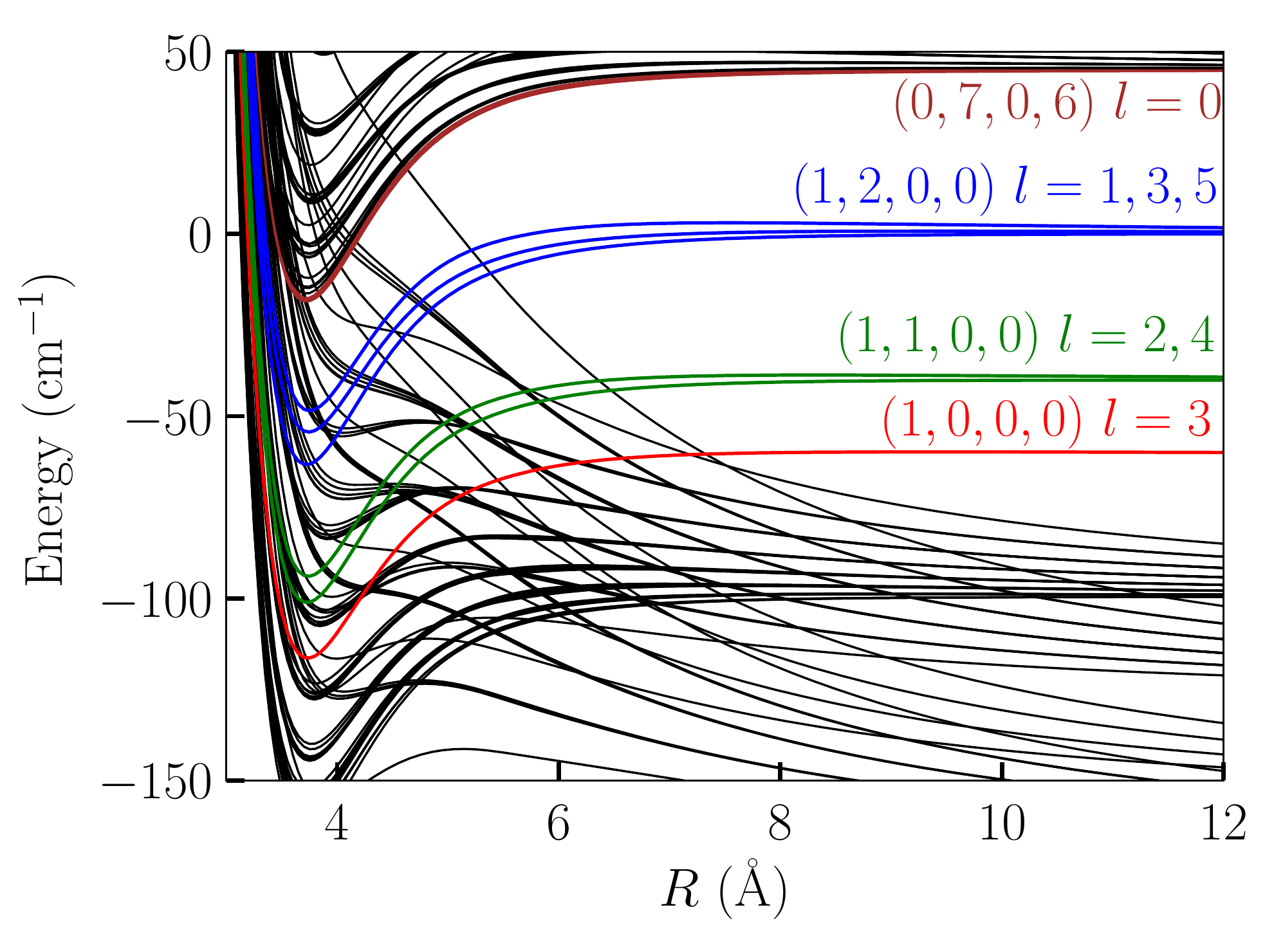}
        \end{center}
      \end{minipage}

    \end{tabular}
    \caption{Adiabatic (left panel) and  diabatic (right panel) potential energy curves for HCl+H$_2$ with the total angular momentum basis set ($J=3$ and $\epsilon_I=-1$). $R$ is the distance between the centers of mass of HCl and H$_2$. Some relevant diabatic states are highlighted with colors and their combined rovibrational states $\alpha$ and associated orbital angular momenta are specified as $\alpha$=($v_1$,$j_1$,$v_2$,$j_2$) and $l$, respectively. The zero energy of the vertical axis is the dissociation limit between HCl ($v_1=1$, $j_1=2$) and H$_2$ ($v_2=0$, $j_2=0$)  which is the initial scattering  state in this study.\\}
    \label{fig_curve}
  \end{center}
\end{figure}

The adiabatic potential energy curves in the left panel of \cref{fig_curve} exhibit multiple avoided crossings, in particular at shorter $R$ range, which are significantly different from the HD+H$_2$ case studied in previously \cite{2018_PRL_Croft}. For HD+H$_2$, the relevant adiabatic potential energy curves did not feature avoided crossings, leading to smooth well-separated curves in the entire range of $R$. Consequently, the collision dynamics of HCl+H$_2$ is much complicated compared to that of HD+H$_2$. 
While the crossings are not necessarily strongly avoided, we observe several series of crossings through a wide $R$ range due to the repulsive curves associated with higher partial waves for the highly rotationally excited molecules in their vibrational ground states.

While the diabatic potential energy curves (right panel) do not contain the information about the strength of coupling with other channels, they are  more easier to interpret. Near the potential minimum in $R$ of the blues curves corresponding to the initial rovibrational state $\alpha\equiv(1,2,0,0)$, we observe some black curves and a brown curve associated with the closed channels such as $\alpha\equiv(0,7,0,6)$ come down in energy below the dissociation limit of initial state (blue curves). The bound states in those closed channels may couple with incoming partial waves in the incident channel  and lead to Feshbach resonances. A  bound state calculation of the {\it s\,}-wave component of $\alpha\equiv(0,7,0,6)$ channel (brown curve) yields  the lowest bound state energy to be about 13.8 cm$^{-1}$ relative to the incident channel. This rules out  a Feshbach resonance  in the cold energy regime. Therefore, the resonances in the cold energy regime described in this work are shape resonances caused by the centrifugal barriers for the incoming partial waves.      
\newline


\begin{center}
	\textbf{\large S-I\hspace{-.1em}I\hspace{-.1em}I.\ Differential cross section}
\end{center}

In the main text we discussed the ICS for rotational quenching of HCl ($v_1=1$, $j_1=2$) and demonstrated the significant stereodynamic effect in the resonance region. The differential cross section (DCS) is generally more sensitive to the preparation of alignment and orientation of initial molecules if examined under a fine resolution of the scattering angle $\theta$. In the following, we explore stereodynamic effect on DCS for the rotational quenching processes explored in the main text.

\begin{figure}[htbp!]
  \begin{center}
    \begin{tabular}{c}

      \begin{minipage}{0.33\hsize}
        \begin{center}
          \includegraphics[ width=6.0cm]{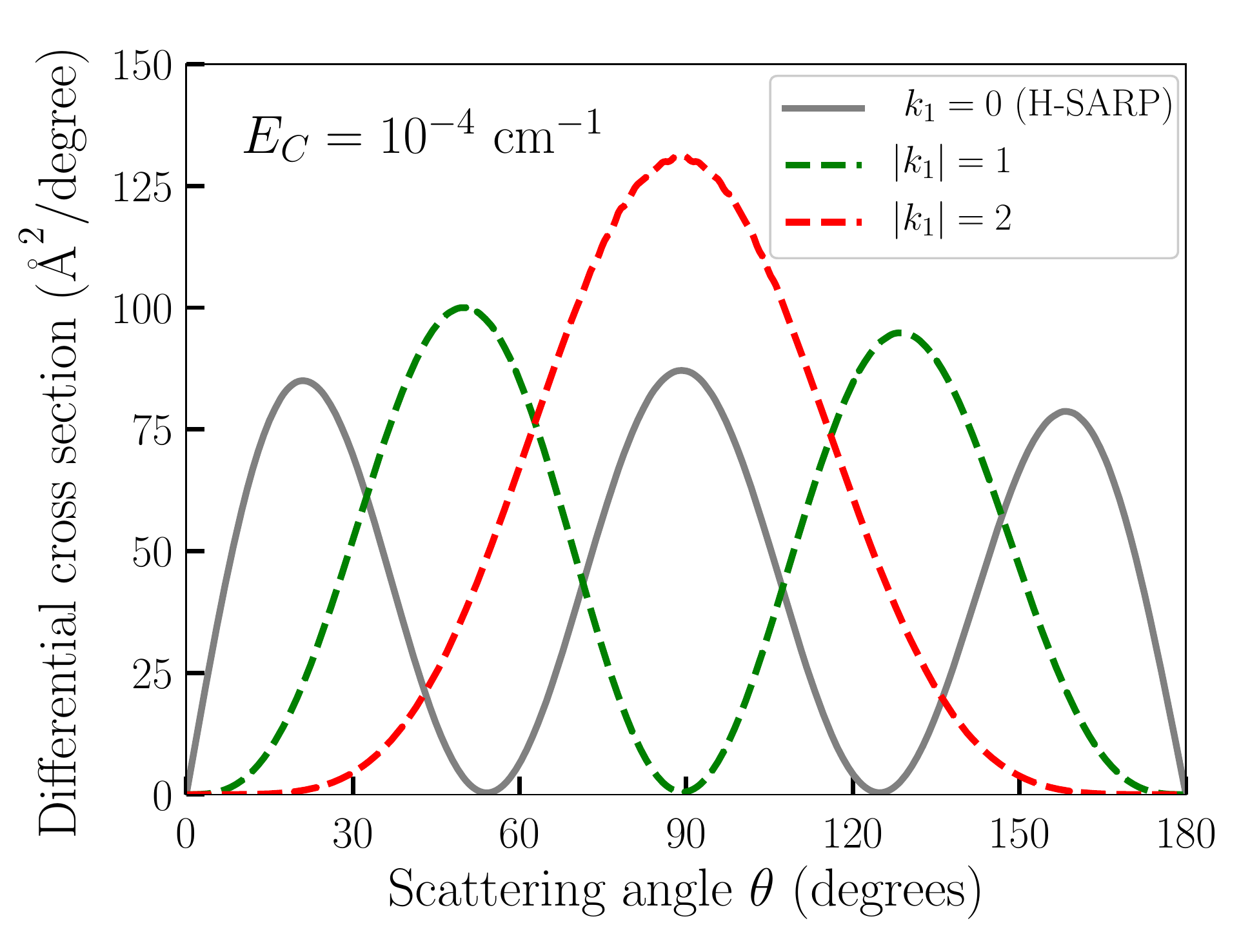}
        \end{center}
      \end{minipage}

      \begin{minipage}{0.33\hsize}
        \begin{center}
          \includegraphics[ width=6.0cm]{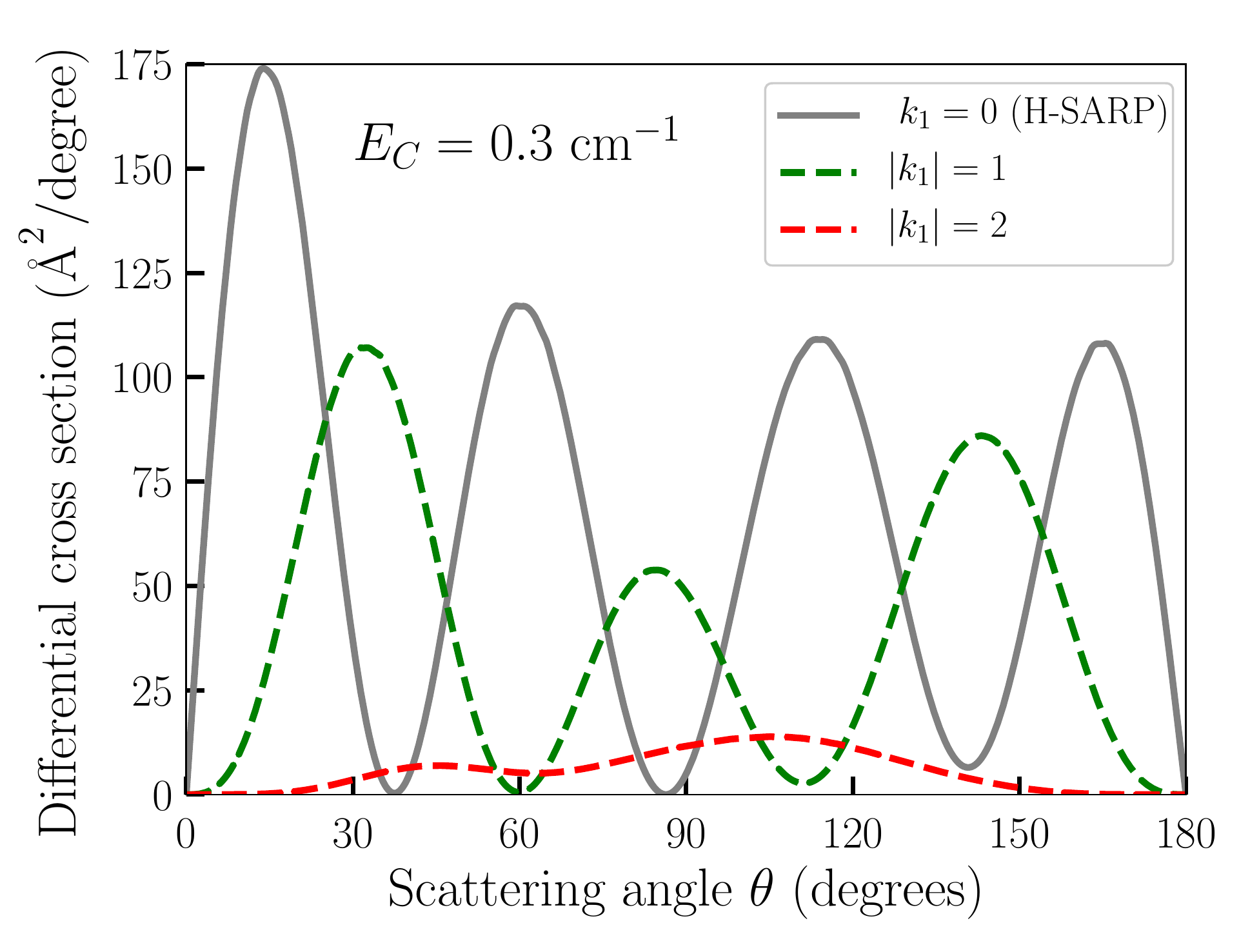}
        \end{center}
      \end{minipage}

      \begin{minipage}{0.33\hsize}
        \begin{center}
          \includegraphics[ width=6.0cm]{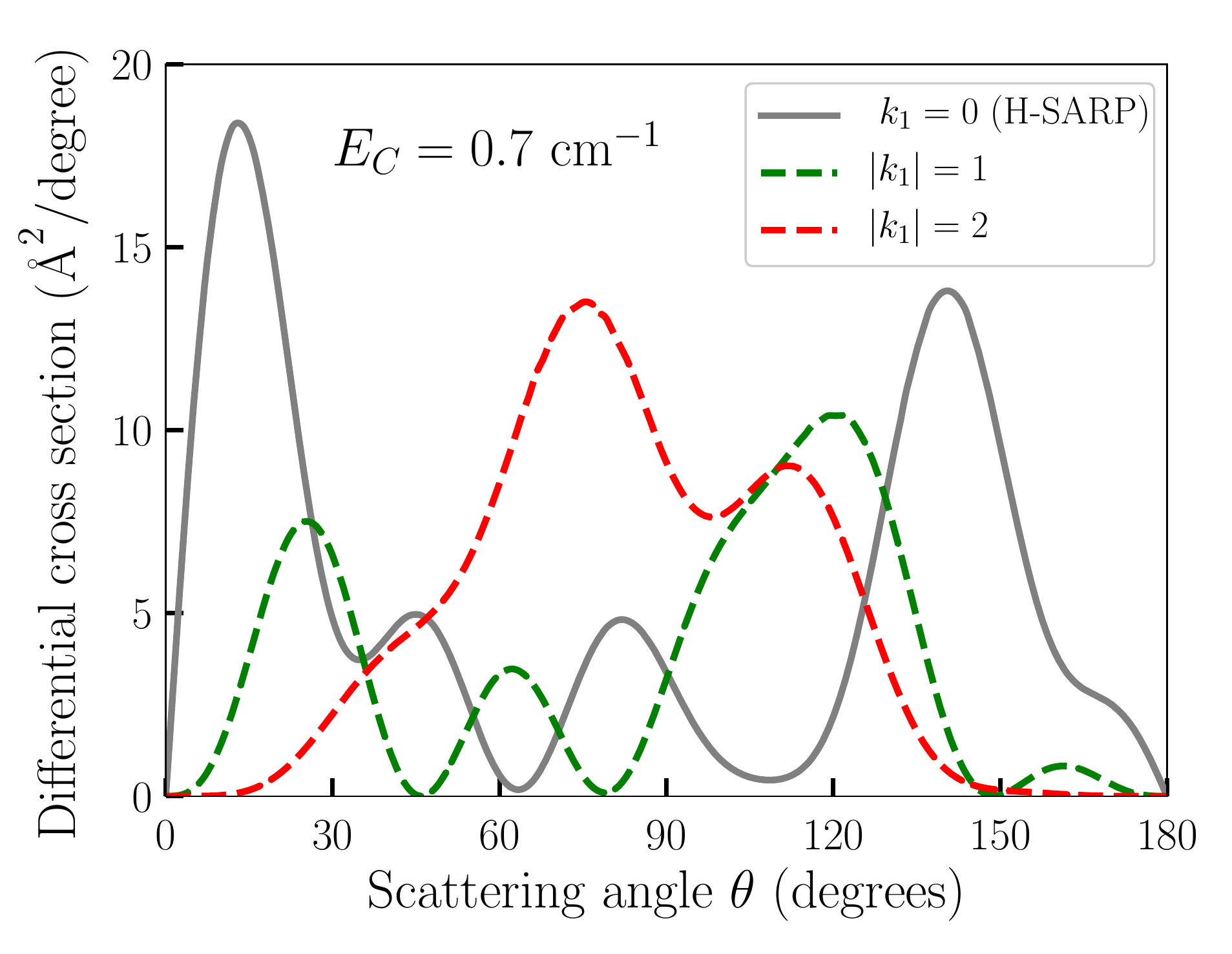}
        \end{center}
      \end{minipage}

    \end{tabular}
    \caption{Differential cross section for the rotational quenching of HCl ($v_1=1$,\ $j_1=2$ $\rightarrow$ $v'_1=1$,\ $j'_1=0$) due to the collision with {\it para}-$\text{H}_2$ ($v_2=0$,\ $j_2=0$). Initial rotational state of HCl has a specific value of the projection quantum number $k_1$. Collision energies are $10^{-4}$ cm$^{-1}$ (left panel), 0.3 cm$^{-1}$ (middle panel) and 0.7 cm$^{-1}$ (right panel).}
    \label{fig_DCS_F10}
  \end{center}
\end{figure}

In previous experiments for HD collisions with H$_2$, D$_2$ and He \cite{2017_Science_Perreault,2018_NatChem_Perreault,SARP_HD-He}, the angular $\theta$ distribution of the rotationally de-excited HD  in a specific rotational state of $j'_1$ was probed by combining multi-photon ionization and time-of-flight spectrum. 
Since the azimuthal angle ($\phi$) dependence of the de-excited HD ($j'_1$) was not observed in the experiments, the $\phi$ dependence is averaged over by taking the integral from 0 to $2\pi$ assuming that there is no detection sensitivity or filtering about $\phi$ in the experiments. 
To simplify the discussion, DCS is calculated with oriented initial HCl rotational state specified by the value of $k_1$ (\cref{fig_DCS_F10}). Like  ICS, the DCS with a general initial HCl alignment by SARP laser can be estimated from the $\beta$-dependence of the weighting factor in \cref{fig_d}.
\newline

\begin{center}
	\textbf{ a). HCl($v_1=1$, $j_1=2$) + H$_2$($v_2=0$, $j_2=0$) \ \,$\to$ \ HCl($v'_1=1$, $j'_1=0$) + H$_2$($v'_2=0$, $j'_2=0$)}
\end{center}

The DCS with the initial state of HCl ($v_1=1$, $j_1=2$, $k_1$) at a collision energy of $E_C=10^{-4}$ cm$^{-1}$ is shown in the left panel of \cref{fig_DCS_F10}. The resulting angular distributions exhibit symmetric behavior about $\theta=\pi/2$ and clear periodic oscillations with respect to $\theta$, indicating that the only the incoming {\it s\,}-wave ($l=0$) 
and the associated outgoing partial wave contribute substantially at such a low energy. The ICS does not have a strong $k_1$ dependence at off-resonant energies, low energies in particular. Thus it is difficult see the steric effect by observing the ICS alone. The $k_1$ dependence of the angular distributions provides a much more sensitive probe of stereodynamics.

At the peak of the ICS ($E_C=0.3$ cm$^{-1}$) due to the incoming {\it p\,}-wave ($l=1$) shape resonance (middle panel of \cref{fig_DCS_F10}), the symmetry of the angular distribution is lost, in particular for $|k_1|=2$ while the periodic oscillations of the cross sections as a function of $\theta$ still remain for $k_1=0$ and $|k_1|=1$. The resonance gives rise to a completely different DCS compared to the off-resonant DCS in the low energy region. 
Furthermore, as shown in the right panel, even at an energy ($E_C=0.7$ cm$^{-1}$) not far from the resonance, the DCS changes significantly. At this energy, the DCS with $k_1=0$ and $|k_1|=2$ show markedly different angular distributions, thus it would be possible to observe dramatic steric effect if we can control the initial population between these $k_1$ components. While DCS provides a much more sensitive prob of steric control, angular distribution would be governed by that given at the peak of the resonance when the peak is isolated and pronounced if we do not have a fine resolution in collision energy.

\begin{center}
	\textbf{ b). HCl($v_1=1$, $j_1=2$) + H$_2$($v_2=0$, $j_2=0$) \ \,$\to$ \ HCl($v'_1=1$, $j'_1=1$) + H$_2$($v'_2=0$, $j'_2=0$)}
\end{center}

\begin{figure}[htbp!]
  \begin{center}
    \begin{tabular}{c}

      \begin{minipage}{0.5\hsize}
        \begin{center}
          \includegraphics[ width=7.0cm]{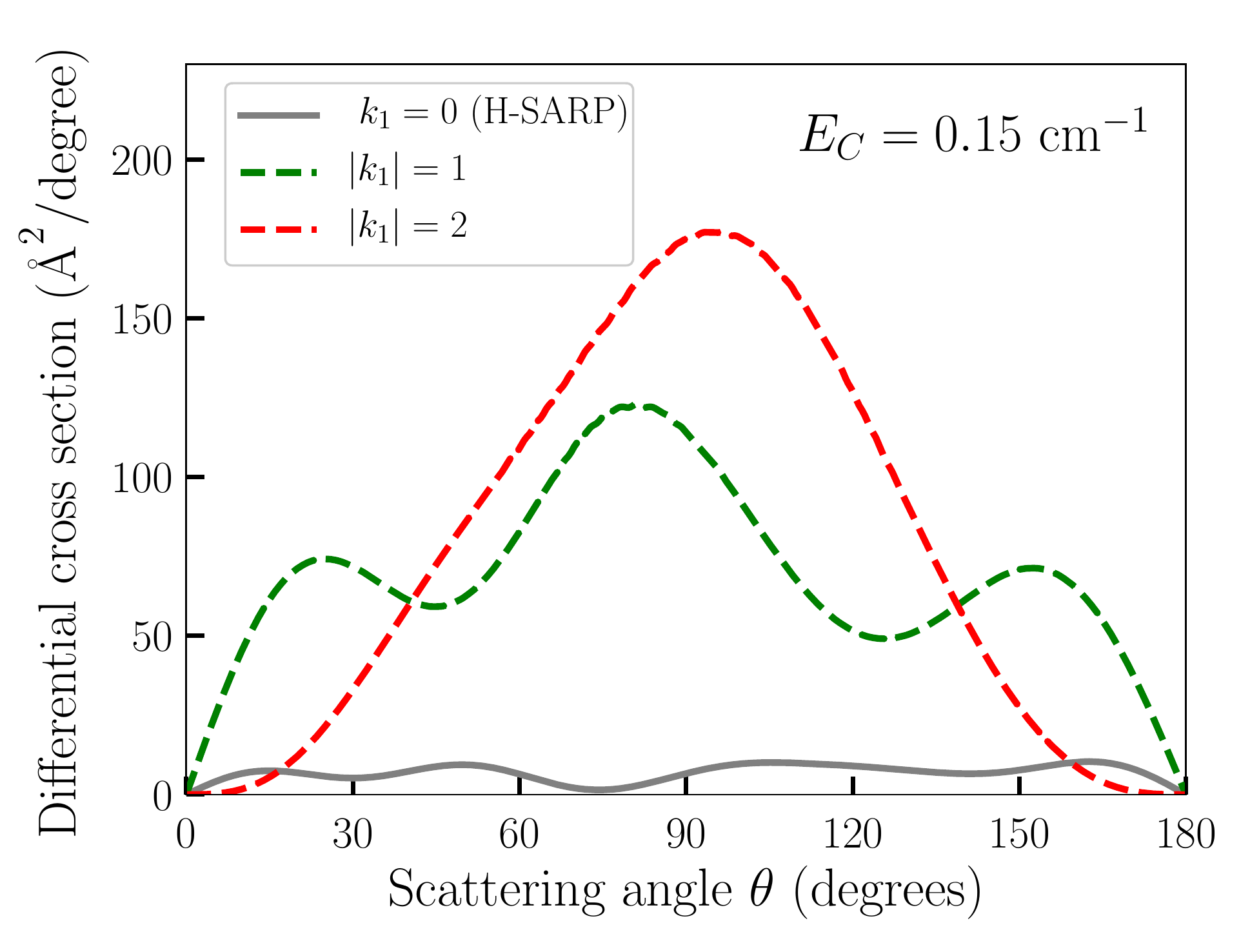}
        \end{center}
      \end{minipage}

      \begin{minipage}{0.5\hsize}
        \begin{center}
          \includegraphics[ width=7.0cm]{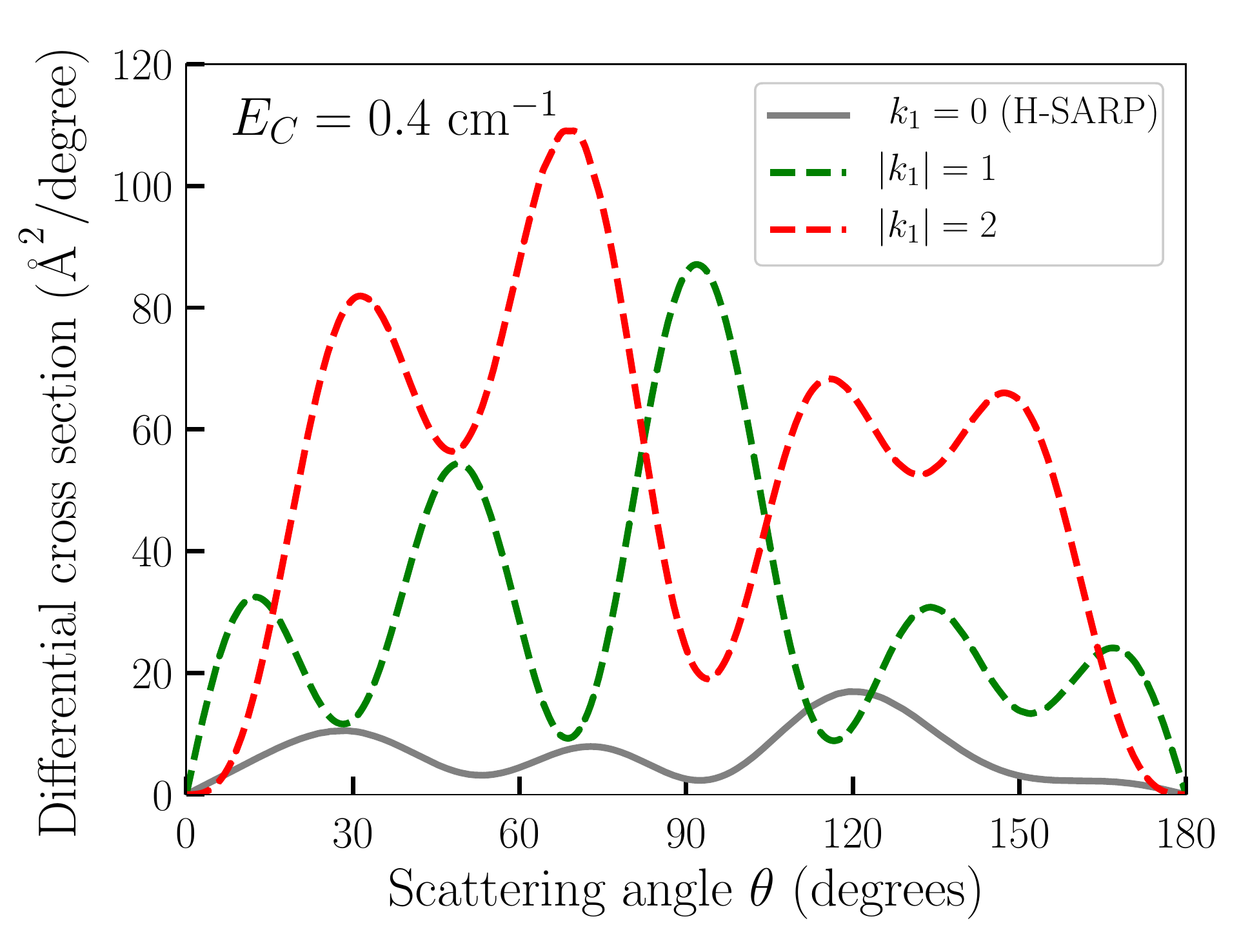}
        \end{center}
      \end{minipage}

    \end{tabular}
\end{center}
\caption{Differential cross section for the rotational quenching of HCl ($v_1=1$,\ $j_1=2$ $\rightarrow$ $v'_1=1$,\ $j'_1=1$) due to the collision with {\it para}-$\text{H}_2$ ($v_2=0$,\ $j_2=0$). Initial rotational state of HCl has a specific value of the projection quantum number $k_1$.  Collision energies are 0.15 cm$^{-1}$ (left panel) and 0.4 cm$^{-1}$ (right panel).
}
\label{fig_I12F11_k_DCS}
\end{figure}

For the final state $j'_1=1$ of the HCl molecule, the initial $k_1$ dependence of the DCS is shown in \cref{fig_I12F11_k_DCS} at two energies corresponding to the prominent peaks of the overlapping resonances. The dramatically different angular dependence  of the DCS at the two energies indicate that the mechanisms of these resonances are completely different. 
at the collision energy, where we observe the peak of the ICS due to the incoming {\it f\,}-wave ($l=3$) resonance, (right panel) shows more oscillatory behavior 
compared to
the incoming {\it p\,}-wave ($l=1$) resonance case (left panel), indicating higher outgoing partial waves ($l'$) are involved 
for the incoming {\it f\,}-wave resonance. 
For both cases, the primary contributions arise from $|k_1|=1$ and 2, a feature also observed in Fig4.~(b) of the main text.

\begin{center}
	\textbf{\large S-I\hspace{-.1em}V.\ Vibrational relaxations}
\end{center}

In the main text, we explored two rotational relaxation processes as in previous studies of HD+H$_2$ within a vibrational manifold of HCl ($v_1=1$). However, if we consider the vibrational relaxation of HCl ($v_1=1 \to v'_1=0$), there are many other energetically allowed  transitions for HCl+H$_2$ even in the cold energy regime. As shown below these transitions are much less efficient than pure rotational quenching. Therefore the effects due to these minor vibrational relaxations are negligible in our present study.  

\begin{figure}[htbp]
  \begin{center}
    \begin{tabular}{c}

      \begin{minipage}{0.5\hsize}
        \begin{center}
          \includegraphics[ width=9.0cm]{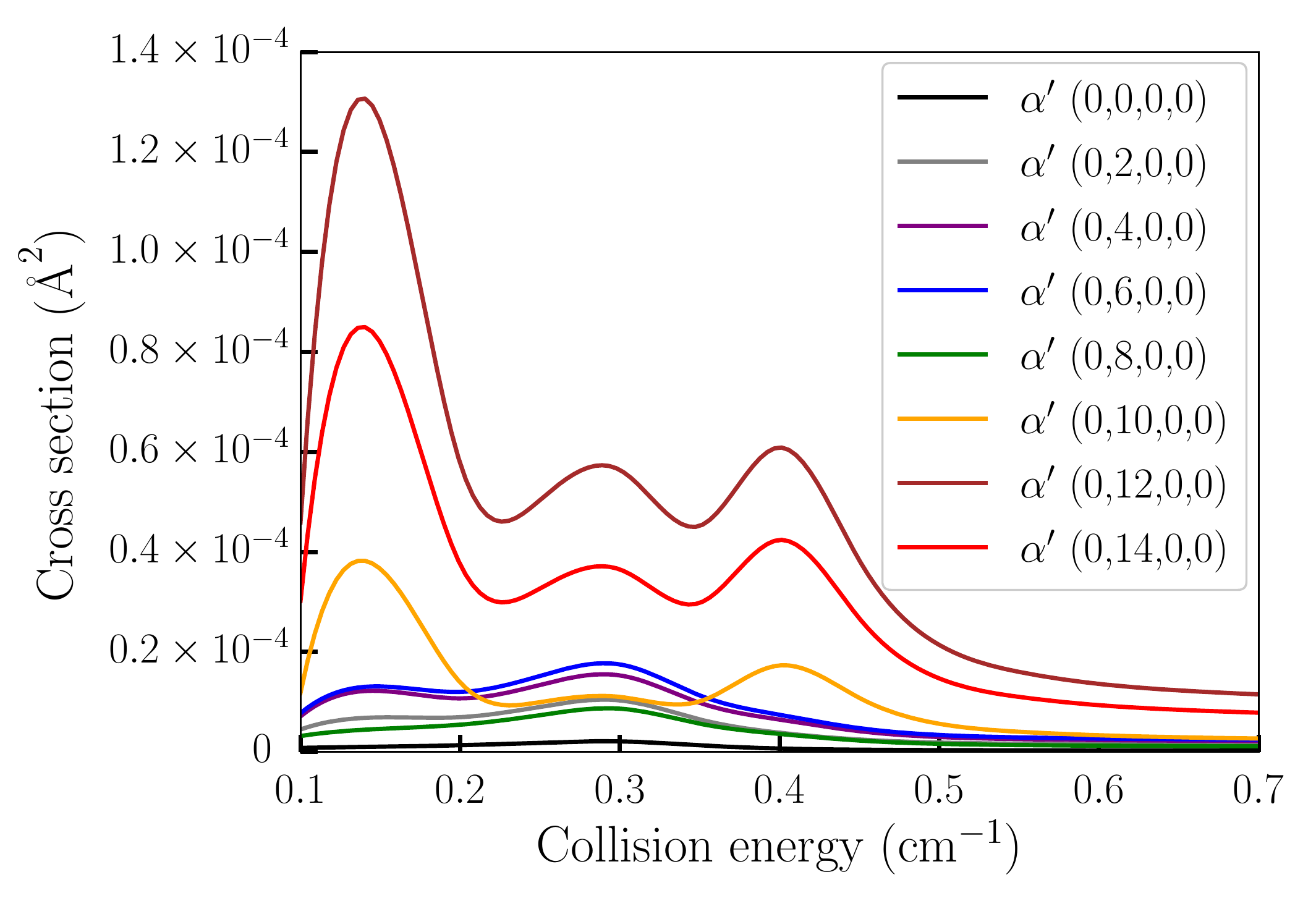}
        \end{center}
      \end{minipage}

      \begin{minipage}{0.5\hsize}
        \begin{center}
          \includegraphics[ width=9.0cm]{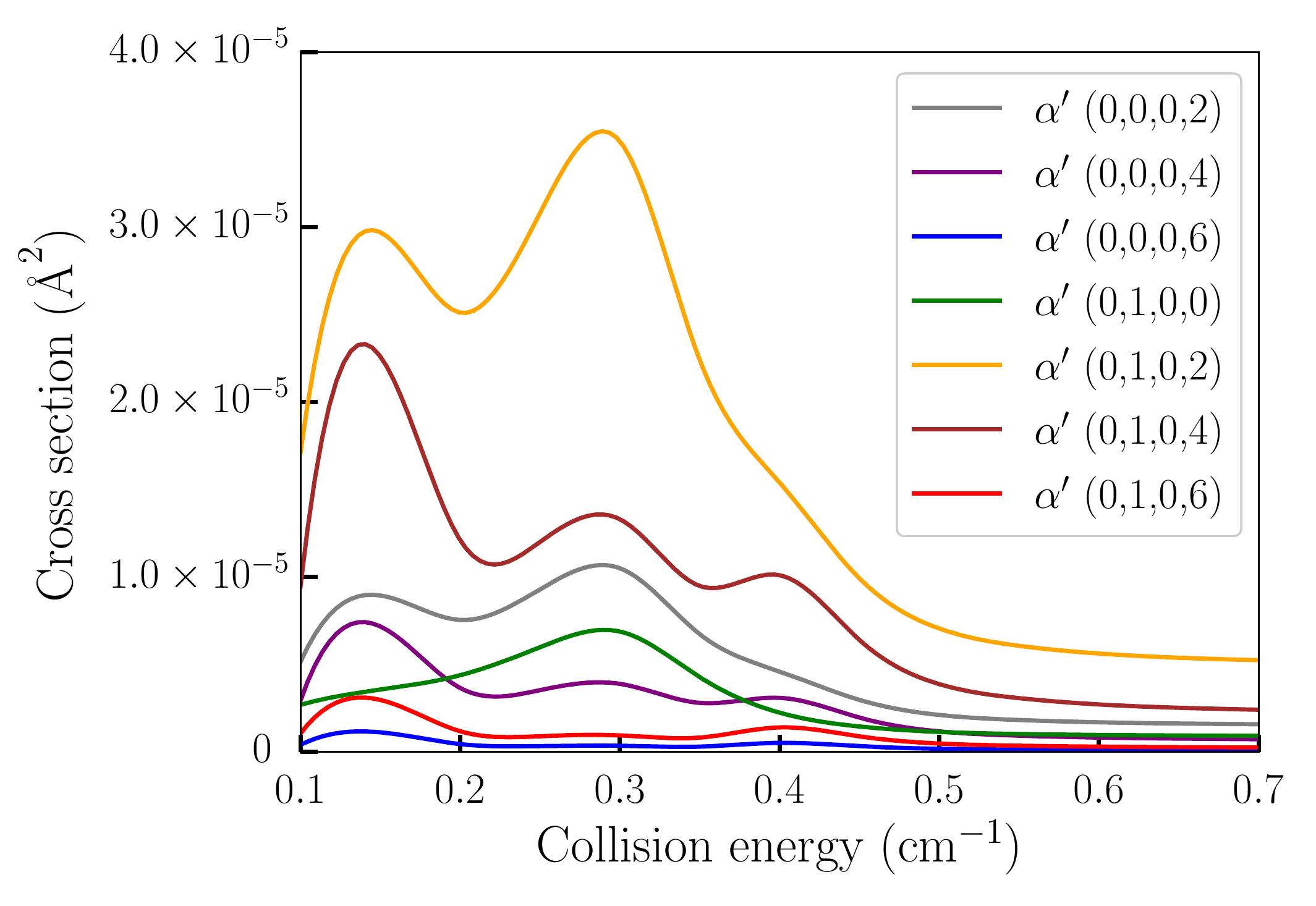}
        \end{center}
      \end{minipage}

    \end{tabular}
    \caption{Integral cross section for the rovibrational transitions of HCl ($v_1=1$,\ $j_1=2$) due to the collision with {\it para}-$\text{H}_2$ ($v_2=0$,\ $j_2=0$) in the absence  of any alignment of molecules. The transitions to the rovibrational states $\alpha'\equiv$ ($v'_1=0$,\ $j'_1$, $v'_2=0$, $j'_2=0$) (left panel) are shown only for even values of $j'_1$ to avoid cluttering. Rotational transitions of {\it para}-H$_2$ ($j_2=0 \to j'_2$) in the transitions to the rovibrational states $\alpha'\equiv $ ($v'_1=0$,\ $j'_1=0/1$, $v'_2=0$, $j'_2$) (right panel) are allowed only to even values of $j'_2$ due to the exchange symmetry of H$_2$.}
    \label{fig_other}
  \end{center}
\end{figure}

There are numerous possible exothermic transition processes from the initial state of HCl ($v_1=1$, $j_1=2$) and H$_2$ ($v_2=0$, $j_2=0$).  One case is the vibrational relaxation of HCl accompanied by a change of its  rotational level. The left panel of \cref{fig_other} shows the integral cross section (ICS) for these transitions for isotropic (no alignment of molecules) collisions, including the pure vibrational relaxation of HCl ($j_1=j'_1=2$, grey curve). To avoid cluttering in the panel, only transitions to even $j'_1$ are displayed. The ICS for vibrational quenching of HCl accompanied by rotational excitation of H$_2$ are displayed in the right panel of \cref{fig_other}. We note that, in this case, only the transitions to even  $j'_2$ are allowed due to the exchange symmetry of H$_2$.   

The presence of the resonance features in the same collision energy regime for these transitions, and also for the rotational quenching studied in the main text, indicate that they are shape resonances associated with specific incoming partial waves and some of them may share the same or similar mechanisms, implying that stereodynamic control of these minor processes would be possible as in the main text. On the other hand, the magnitude and the resonance features are sensitive to the final states, indicating the coupling strength and interference pattern between the multiple resonances strongly depend on the final state.

\end{document}